\documentclass[12pt]{article}
\usepackage{graphicx}
\usepackage{amsfonts}
\usepackage{amssymb,amsmath}
\usepackage{latexsym}
\usepackage{color}
\usepackage{cite}
\usepackage{bm}   
\input{colordvi.tex}

\begin{document}

\begin{titlepage}

\begin{flushright}
IPMU 11-0038 \\
ICRR-Report-582-2010-15
\end{flushright}

\vskip .1in

\begin{center}

{\Large \bf  
Differentiating CDM and Baryon Isocurvature Models with
21 cm Fluctuations
}

\vskip .45in

{\large
Masahiro Kawasaki$^{1,2}$, 
Toyokazu Sekiguchi$^{1,3}$ and 
Tomo Takahashi$^4$ 
}

\vskip .45in

{\em
$^1$Institute for Cosmic Ray Research, University of Tokyo,
Kashiwa 277-8582, Japan  \vspace{0.2cm} \\
$^2$Institute for the Physics and Mathematics of the Universe,
University of Tokyo, Kashiwa, Chiba, 277-8568, Japan \vspace{0.2cm}\\
$^3$Department of Physics and Astrophysics, Nagoya University, 
Nagoya 464-8602, Japan \vspace{0.2cm} \\
$^4$Department of Physics, Saga University, Saga 840-8502, Japan 
}

\end{center}

\vskip .4in

\begin{abstract}

We discuss how one can discriminate models with cold dark matter (CDM)
 and baryon 
isocurvature fluctuations. Although current observations such as cosmic 
microwave background (CMB) can severely constrain the fraction of such isocurvature 
modes in the total density fluctuations, CMB cannot  differentiate 
CDM and baryon ones by the shapes of their power spectra.
However, the evolution of CDM and baryon density fluctuations  
are different for each model, thus it would be possible to discriminate those 
isocurvature modes by extracting information on the fluctuations of CDM/baryon itself. 
We discuss that observations of 21 cm fluctuations can in principle differentiate these modes and 
demonstrate to what extent we can distinguish them with future 
21 cm surveys. We show that, when the isocurvature mode has 
a large blue-tilted initial spectrum, 21 cm surveys can clearly probe the difference.

\end{abstract}

\end{titlepage}

\setcounter{page}{1}

\section{Introduction}

Current cosmological observations are now very precise to give 
a stringent constraint on the adiabaticity of density fluctuations. 
However, some contribution of non-adiabatic fluctuations, so-called isocurvature fluctuations,
 is still allowed (see e.g., \cite{Komatsu:2010fb}).
Isocurvature fluctuations can be generated in cold dark matter (CDM) and 
baryon sectors, 
from which we can have some insight on the nature of dark matter and/or
the mechanism of the baryogenesis. This is because isocurvature fluctuations 
might be produced depending on how dark matter/baryon number is generated. 
Such examples are axion \cite{Axenides:1983hj,Seckel:1985tj,Linde:1985yf,
Linde:1990yj,Turner:1990uz,Linde:1991km,Lyth:1991ub} and
Affleck-Dine baryogenesis mechanism \cite{Enqvist:1998pf,Enqvist:1999hv,
Kawasaki:2001in,Kasuya:2009up} in which 
CDM and baryon isocurvature fluctuations can respectively be generated\footnote{
Other examples producing CDM/baryon isocurvature fluctuations include the curvaton 
model \cite{Enqvist:2001zp,Lyth:2001nq,Moroi:2001ct}, in which the isocurvature 
fluctuations partially/totally correlated with adiabatic ones depending on how 
CDM/baryon are generated \cite{Moroi:2002rd,Lyth:2003ip}. 
Furthermore, isocurvature fluctuations can be generated in neutrino density fluctuations 
and neutrino  velocity  \cite{Bucher:1999re}.
However, we focus on uncorrelated CDM and baryon isocurvature modes in this paper.
}.
If such a mode is found to be non-zero, it would give invaluable 
information on CDM/baryogenesis scenarios. 

Currently, stringent constraints on isocurvature fluctuations come 
from cosmic microwave background (CMB) observations such as WMAP \cite{Komatsu:2010fb}. 
However, CMB cannot discriminate CDM and baryon ones since 
these two isocurvature modes give indistinguishable CMB angular power spectra.
The initial non-zero isocurvature mode of CDM/baryon affects 
fluctuations of photons, at which we are looking in CMB observations, 
only through gravity. CDM and baryon can be regarded as the same 
component as far as the gravity is concerned, hence the effects of 
isocurvature fluctuations arise in exactly the same way for CMB anisotropy
(except the amplitude which depends on the amount of CDM and baryon).

Thus even if we see some signature of existence of the isocurvature mode
from CMB measurements, we cannot tell whether CDM or baryon  is responsible for that.
From the viewpoint of testing dark matter models/baryogenesis mechanism, 
it is very crucial to differentiate these two modes in some way, which is the 
issue we are going to discuss in this paper. 
Although CMB measurement cannot do this task, the evolution of density 
fluctuations of CDM and baryon themselves are not the same for models with 
initial CDM and baryon isocurvature fluctuations. 
Thus we may have some possibility for differentiating those by looking at 
(CDM/baryon) fluctuations directly.

In this paper, we discuss how one can probe the difference between these two 
isocurvature modes by using observations of fluctuations in redshifted 21 cm line absorption.
As mentioned above, CDM and baryon fluctuations evolve differently
when one assume different isocurvature modes (i.e., CDM and baryon ones).
Hence,  if either one can be 
probed directly in some way, we can in principle see the difference. 
Since 21 cm fluctuations effectively probe baryon density fluctuations, the observations of them
can be utilized for this purpose\footnote{
The possibility of probing baryon and CDM fluctuations separately using 21 cm fluctuations was 
discussed in \cite{Barkana:2005xu}. The issue of 
distinguishing baryon and CDM isocurvature modes with 21 cm surveys 
was investigated for  the ``compensated isocurvature 
mode" in  \cite{Gordon:2009wx}. 
See also \cite{Grin:2011tf}.
}.

The organization of this paper is as follows. In the next section, 
we discuss the evolution and effects of density fluctuations of CDM and baryon 
in models with isocurvature fluctuations. We explain in some detail 
what aspect  differs between CDM and baryon isocurvature modes 
by looking at density fluctuations of CDM and baryon in these modes.
Then in Section 3, we show that 21 cm fluctuations can in principle 
differentiate the CDM and baryon isocurvature modes, which is impossible
with CMB observations. We also demonstrate a future constraint on CDM and baryon 
isocurvature  fluctuations by utilizing future 21 cm survey and discuss 
to what extent they can be discriminated.
The final section is devoted to the summary of this paper.

\section{Evolutions of density fluctuations in CDM and baryon 
isocurvature models}

\subsection{Notations}

Before we start the discussion on the evolution of density fluctuations, here we give some 
definitions, in particular, regarding isocurvature fluctuations to set our notation. 

Gauge-invariant isocurvature fluctuations between the components $i$ and $j$ are defined as
\begin{equation}
S_{ij} = -3 H \left( \frac{\delta \rho_i }{\dot{\rho}_i}  - \frac{\delta \rho_j }{\dot{\rho}_j} \right).
\end{equation}
In the following, we mainly consider isocurvature fluctuations of CDM and baryon with respect to radiation,
which we write as
\begin{equation}
S_c  = \delta_c   - \frac{3}{4} \delta_\gamma, \qquad
S_b  = \delta_b   - \frac{3}{4} \delta_\gamma. \qquad
\end{equation}
Furthermore, in this paper we only consider uncorrelated type of isocurvature 
fluctuations. Thus CDM and baryon isocurvature modes can be respectively 
identified as the ones with non-vanishing $S_c = \delta_c (\tau \rightarrow 0)$ 
and $S_b = \delta_b (\tau \rightarrow 0)$, with $\tau$ being the conformal time,
 at the early times when the initial conditions are set for 
cosmological density fluctuations. 
Then the primordial power spectrum for isocurvature fluctuations 
is defined as 
\begin{eqnarray}
\mathcal{P}_{S_i}(k)(2\pi)^3 \delta^{(3)} (\bm{k} - \bm{k}') 
= \frac{k^3}{2\pi^2} \left\langle S_i (\bm{k})  S_i (\bm{k}')^{\ast} \right\rangle,
\end{eqnarray}
where $i=c$ and $b$ indicating CDM and baryon, respectively,  and we parametrize $\mathcal{P}_{S_i}$ as 
\begin{equation}
\mathcal{P}_{S_i} (k) = \mathcal{P}_{S_i} (k_0) \left( \frac{k}{k_0} \right)^{n_s^{(i)}-1}
\end{equation}
with $n_s^{(i)}$ is the spectral index for the mode $i$. 
In this paper, we take the reference scale $k_0$ as $k_0=0.002~{\rm Mpc}^{-1}$.
For CDM and baryon isocurvature modes,
we denote it as $n_s^{\rm (CDMiso)}$ and $n_s^{\rm (biso)}$, respectively.
For the adiabatic (curvature) fluctuations,  similarly to the above, one usually defines  as
\begin{eqnarray}
\mathcal{P}_{\zeta} (k)(2\pi)^3 \delta^{(3)} (\bm{k} - \bm{k}') 
= \frac{k^3}{2\pi^2} \left\langle \zeta (\bm{k}) \zeta (\bm{k}')^{\ast} \right\rangle,
\end{eqnarray}
where $\zeta$ is the curvature perturbation and the power spectrum $\mathcal{P}_{\zeta}$
can also be written as
\begin{equation}
\mathcal{P}_{\zeta} (k) = \mathcal{P}_{\zeta} (k_0) \left( \frac{k}{k_0} \right)^{n_s-1},
\end{equation}
with $n_s$ being the spectral index for the adiabatic mode.

Current cosmological observations such as WMAP indicate that 
cosmic density fluctuations are almost adiabatic and 
pure isocurvature fluctuations of any modes are excluded.  However, 
some fraction (contamination) of isocurvature fluctuations are still allowed. 
To characterize the fractions of the isocurvature modes, we define the quantities $r_a$ and $r_b$ 
as\footnote{
In some literature such as \cite{Komatsu:2010fb,Bean:2006qz}, 
the fraction of isocurvature fluctuations 
is defined as
\begin{equation}
\alpha = \frac{\mathcal{P}_{S_c} (k_0)}{\mathcal{P}_\zeta (k_0) + \mathcal{P}_{S_c} (k_0)},
\end{equation}
which is a bit different from the one used here.  
However, as far as the fraction of the isocurvature 
fluctuations is small, both definitions are almost equivalent.
}
\begin{equation}
r_c \equiv \frac{\mathcal{P}_{S_c} (k_0)}{\mathcal{P}_\zeta (k_0)}, \qquad
r_b \equiv \frac{\mathcal{P}_{S_b} (k_0)}{\mathcal{P}_\zeta (k_0)}.
\end{equation}
From current observations, $r_c$ is constrained to be less than 
about 10 \% \cite{Komatsu:2010fb}.
(The corresponding constraint on $r_b$ becomes less severe by the factor 
of $(\Omega_c/\Omega_b)^2$.) Thus we use $r_c=0.1$ as a 
reference value unless otherwise stated in the following.

\subsection{Evolutions of density fluctuations}
\label{subsec:evolution}

Now we study the evolutions of 
density fluctuations in models with CDM and baryon isocurvature fluctuations and 
see the difference of those between in these isocurvature modes, which could be 
probed with 21 cm fluctuations. In addition, we also discuss why CMB cannot discriminate
these two isocurvature modes at linear perturbation level. 

In the following, we investigate density perturbations in the synchronous gauge 
where the metric perturbations  are defined as
\begin{equation}
ds^2 = a^2 (\tau) [ -d\tau^2 + (\delta_{ij} + h_{ij} ) dx^i dx^j ], 
\end{equation}
with 
\begin{equation}
h_{ij} ({\bm x}, \tau) = \int dk^3 e^{ i {\bm k}\cdot {\bm x}} 
\left[ 
\hat{\bm k}_i \hat{\bm k}_j h({\bm k},\tau) 
+  \left( \hat{\bm k}_i \hat{\bm k}_j  - \frac{1}{3} \delta_{ij} \right) 
6 \eta ({\bm k}, \tau)
\right]. 
\end{equation}
Here $\tau$ is the conformal time and $a$ is the scale factor.
We denote density fluctuations of a species $i$ as 
$\delta_i \equiv \delta \rho_i / \bar{\rho}_i$.
In particular, here we study  those of CDM $\delta_c$, 
baryon $\delta_b$ and photons  $\delta_\gamma$. 
We follow the notation of \cite{Ma:1995ey}.

In this gauge, the perturbed Einstein equations are:
\begin{eqnarray}
k^2 \eta - \frac{1}{2} \mathcal{H} \dot{h}  & = & - 4 \pi G a^2 \delta \rho, 
\label{eq:doth}\\
k^2 \dot{\eta}  & = & - 4 \pi G a^2 (\bar{\rho} + \bar{P} ) \theta, 
\label{eq:doteta}\\
\ddot{h} + 2 \mathcal{H} \dot{h} - 2 k^2 \eta &=& - 24\pi G a^2 \delta P, \\
\ddot{h} + 6 \ddot{\eta} + 2 \mathcal{H} \left( \dot{h} + 6 \dot{\eta} \right) 
- 2 k^2 \eta &=& - 24\pi G a^2 (\bar{\rho} + \bar{P} ) \sigma, 
\end{eqnarray}
where a dot represents the derivative with respect to the conformal 
time $\tau$ and $\mathcal{H} = \dot{a}/{a}$ is the Hubble parameter 
defined with the conformal time derivative.  
$\delta \rho$ and $\delta P$ are fluctuations of energy density and pressure for the total component.
$\theta$ is a variable defined as $(\bar{\rho} + \bar{P})\theta = \sum_i (\bar{\rho}_i + \bar{P}_i) i k v_i$ 
with $v_i$ being velocity perturbation 
of species $i$ and $\sigma$ represents the anisotropic stress of 
the total component. 
The equation for the CDM fluctuations is
\begin{equation}
\label{eq:deltac}
\dot{\delta}_c  = - \frac{1}{2} \dot{h}. 
\end{equation}
Here we note that the velocity perturbation for CDM is set as $\theta_c = 0$ so 
that we can fix the gauge freedom in the synchronous gauge. 
For baryon, we have
\begin{equation}
\dot{\delta}_b  = - \theta_b- \frac{1}{2} \dot{h}, \qquad \qquad
\dot{\theta}_b = \mathcal{H} \theta_b 
+ R a  n_e \sigma_T (\theta_\gamma  - \theta_b),
\label{eq:theta_b}
\end{equation}
where $\sigma_T$ is the cross section for the Thomson 
scattering and $n_e$ is the number density of electron, 
and we also have defined $R \equiv 4 \bar{\rho}_\gamma / (3\bar{\rho}_b )$. 
For photons, the equations are:
\begin{equation}
\dot{\delta}_\gamma  
= - \frac{4}{3} \theta_\gamma - \frac{2}{3} \dot{h}, \qquad \qquad
\dot{\theta}_\gamma  = k^2 \left( \frac{1}{4} \delta_\gamma - \sigma_\gamma \right) 
- a  n_e \sigma_T (\theta_\gamma  - \theta_b).
\label{eq:theta_g}
\end{equation}
Similarly, for neutrinos, the equations of motion are
\begin{equation}
\dot{\delta}_\nu  
= - \frac{4}{3} \theta_\nu - \frac{2}{3} \dot{h}, \qquad \qquad
\dot{\theta}_\nu  = k^2 \left( \frac{1}{4} \delta_\nu - \sigma_\nu \right).
\label{eq:deltanu}
\end{equation}
Here we have omitted the equation for the anisotropic stress 
and higher multipole moments for photons and those for 
neutrinos since they are not so relevant to our discussion and 
we do not discuss in the following.

During the epoch deep in the radiation dominated era, 
the time scale of the Thomson scattering is much shorter than the 
Hubble expansion, which dampens  anisotropic stress of photons and 
drives the velocity perturbations of baryon and photons the same. 
Thus in the above equations, we can set $\sigma_\gamma =0$ and $\theta_b = \theta_\gamma$
 in early times when the initial conditions are given. 
Furthermore, on superhorizon scales, the terms with velocity perturbations can be regarded as 
smaller by a factor of $k$\footnote{
This is not true in the neutrino density and velocity isocurvature modes 
\cite{Bucher:1999re}, which we do not consider in this paper.
}, which allows us to neglect those terms at leading order of $k$.
Then we obtain the following relations at the leading order in $k\tau$:
\begin{equation}
\label{eq:delta_diff}
\dot{h} = -2 \dot{\delta}_c = -2 \dot{\delta}_b = -\frac{3}{2} \dot{\delta}_\gamma,
\end{equation}
which shows that the evolutions of $\delta_i$ follows that of $h$ with some numerical 
factors except a constant term which can be determined by the initial condition for 
the perturbations.

On the other hand, inside the horizon, the evolutions of fluctuations are different from 
those on superhorizon scales.
Although CDM holds the relation with the metric perturbation $h$ with Eq.~\eqref{eq:deltac}, 
on subhorizon scales where $k \tau \gg 1$, baryon and photons obey the following equations:
\begin{eqnarray}
\label{eq:deltab_sub}
\ddot{\delta}_b + \mathcal{H} \dot{\delta}_b + c_s^2 k^2 \delta_b
& = & 
- \frac{1}{2} \mathcal{H} \dot{h} - \frac{\ddot{h}}{2}, \\
\label{eq:deltag_sub}
\ddot{\delta}_\gamma + \frac{1}{3}k^2 \dot{\delta}_\gamma 
 & = &  
 - \frac{2}{3} \ddot{h}.
\end{eqnarray}
Since inside the horizon, $\delta_c$ grows as $\delta_c \propto \ln a$ during radiation-dominated (RD) 
era and $\delta_c \propto a$ during matter-dominated (MD) era, the metric perturbation $h$ also 
evolves as $\dot{h} \propto 1/a$ and $a^{1/2}$ during RD and MD eras, respectively. 
Since the scale factor can be given by the conformal time $\tau$ as $a \propto \tau$ and $\tau^2$ 
during RD and MD eras, respectively,   from Eqs.~\eqref{eq:deltab_sub} and \eqref{eq:deltag_sub}, 
$\delta_b$ and $\delta_\gamma$ show oscillatory behaviors with 
almost constant amplitudes except the epoch of the transition from 
RD to MD. As will be discussed shortly, these behaviors are the same 
for adiabatic and isocurvature modes except the phases of the oscillations due to the difference 
of the initial conditions of these modes which are given when the fluctuations enter the horizon.

\begin{figure}
  \begin{center}
      \resizebox{80mm}{!}{\includegraphics{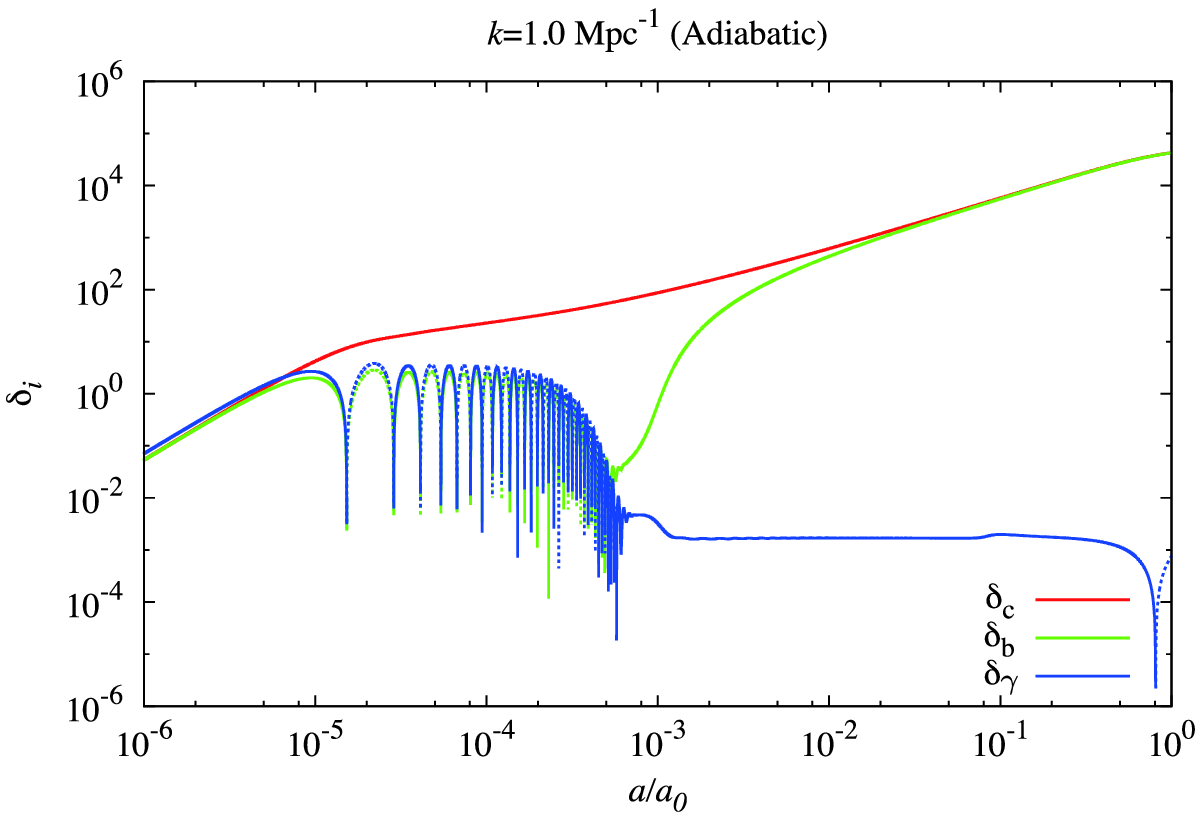}} 
      \resizebox{80mm}{!}{\includegraphics{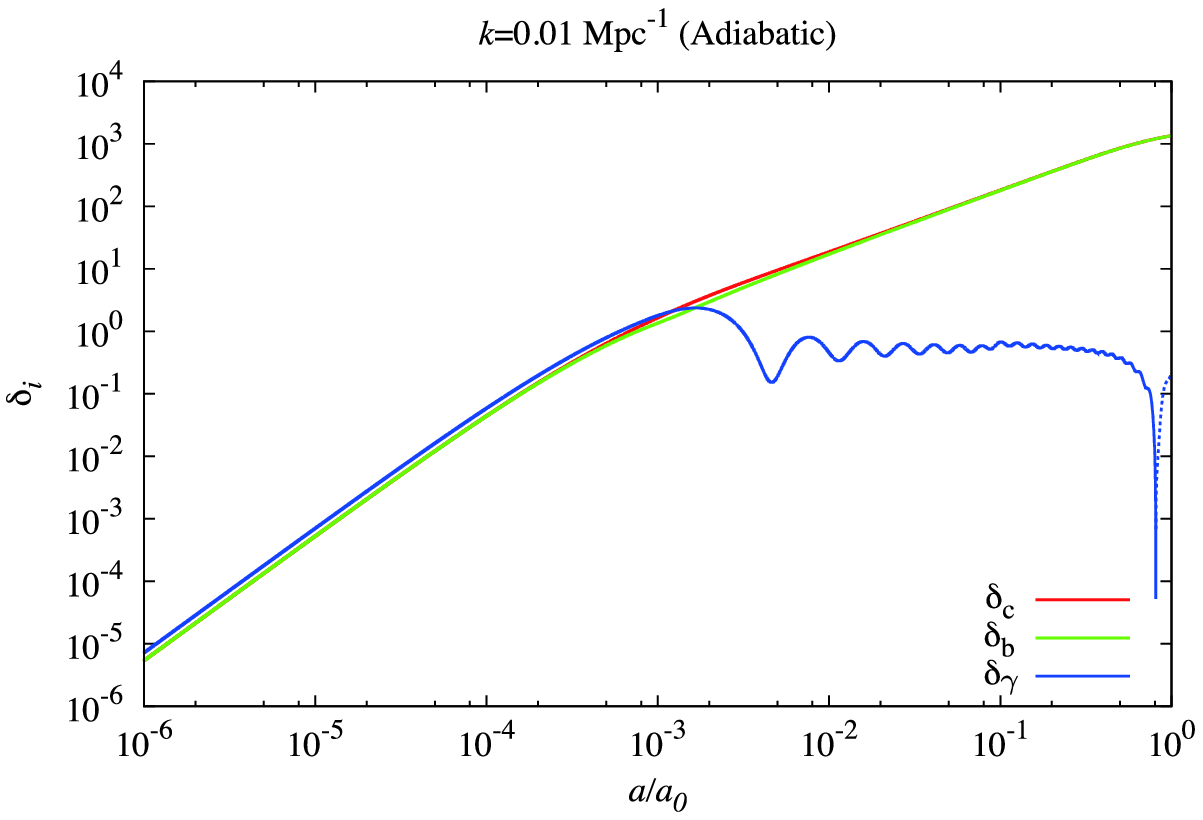}}
      \resizebox{80mm}{!}{\includegraphics{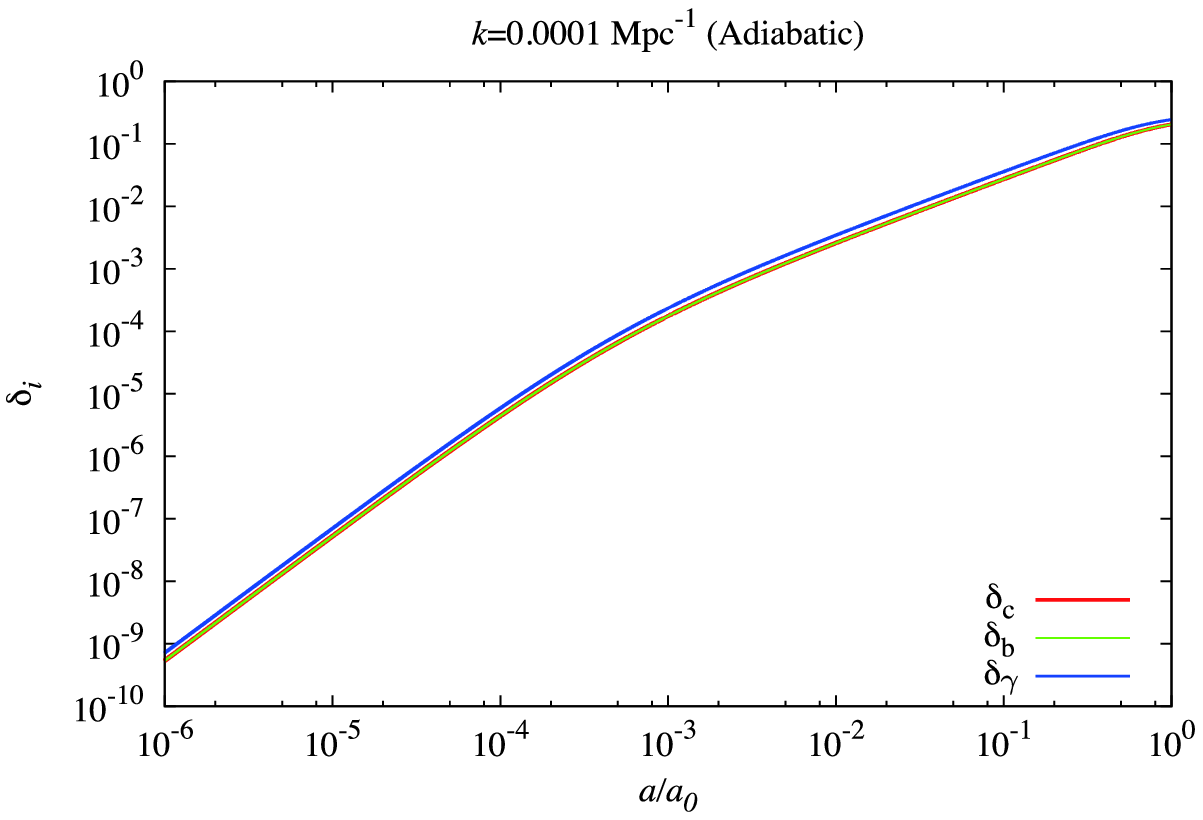}} 
  \end{center}
  \caption{Evolutions of $\delta_c, \delta_b$ and  $\delta_\gamma$ for adiabatic mode 
  in the synchronous gauge. Here we show those for $k=1.0$ (top), $0.01$ (middle)
  and $0.0001~\mathrm{Mpc}^{-1}$ (bottom).
  }
   \label{fig:evolv_adi}
\end{figure}

\begin{figure}
  \begin{center}
    \begin{tabular}{cc}
      \resizebox{80mm}{!}{\includegraphics{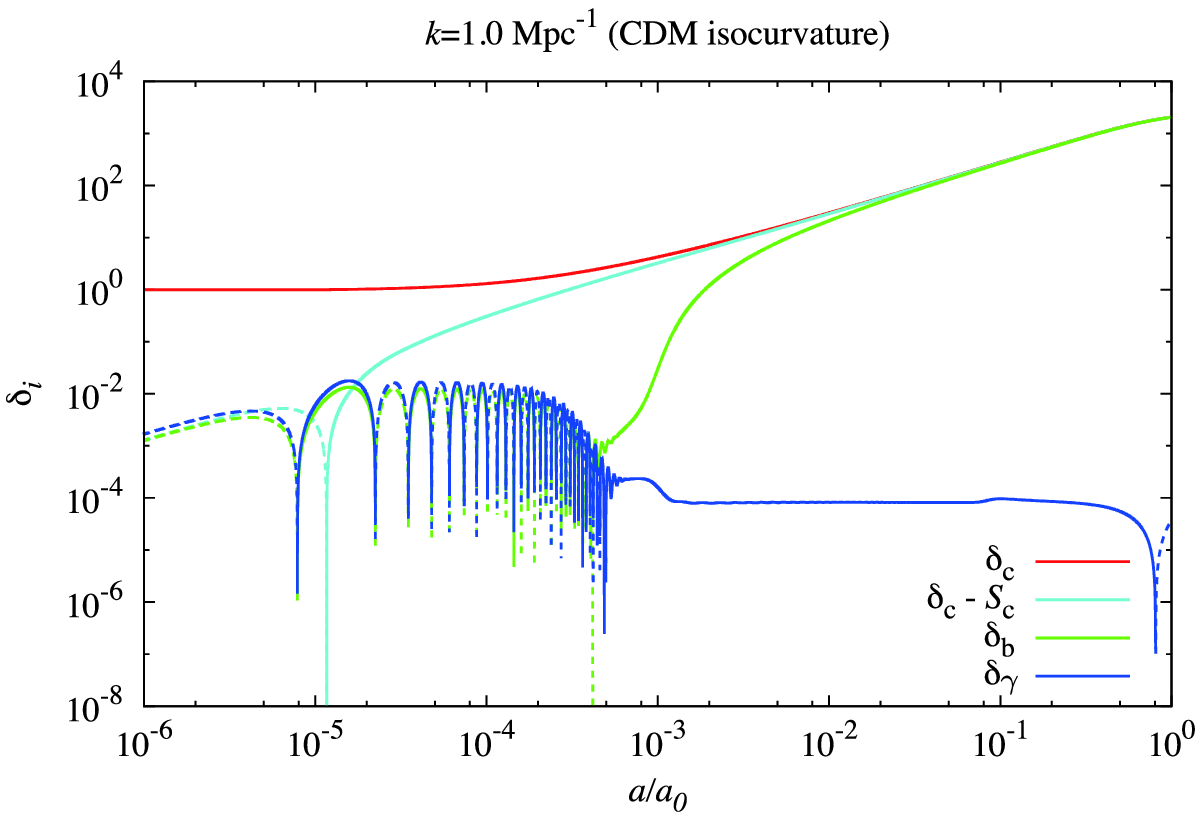}} &
      \hspace{0mm}
      \resizebox{80mm}{!}{\includegraphics{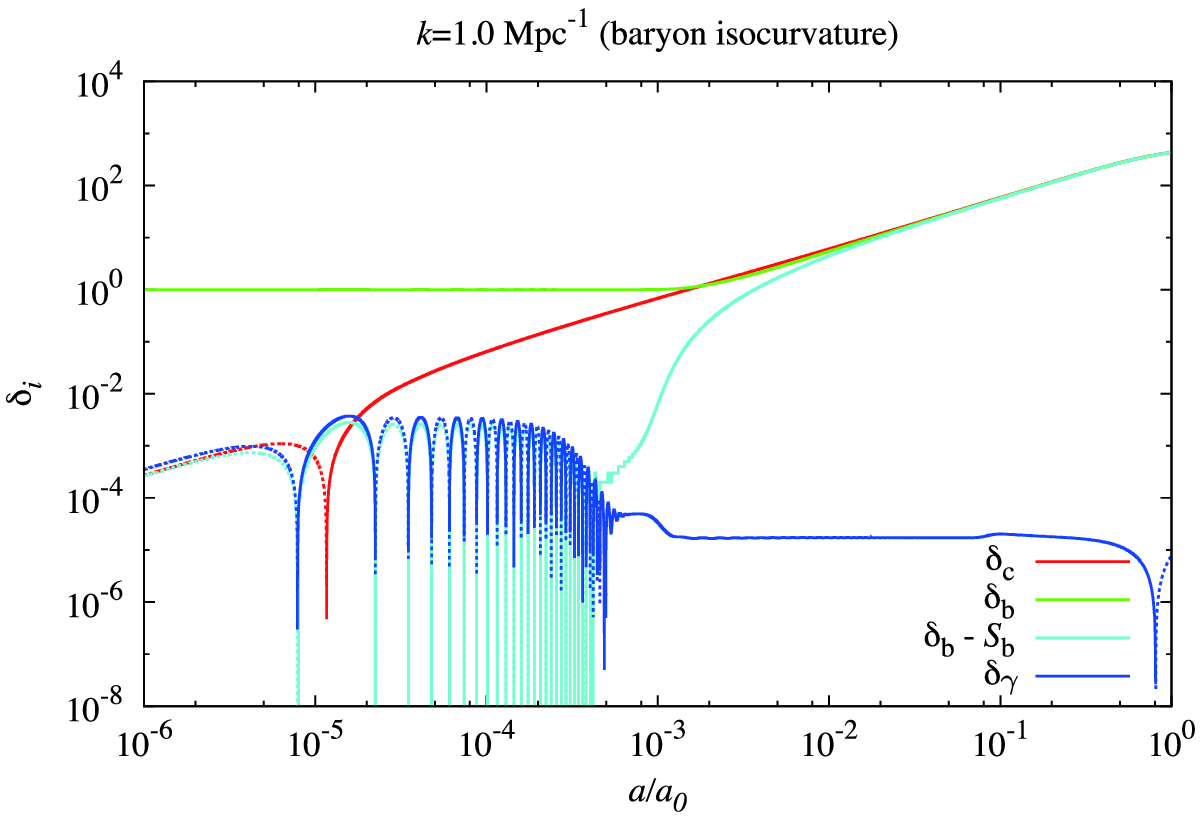}}  
     \\
      \resizebox{80mm}{!}{\includegraphics{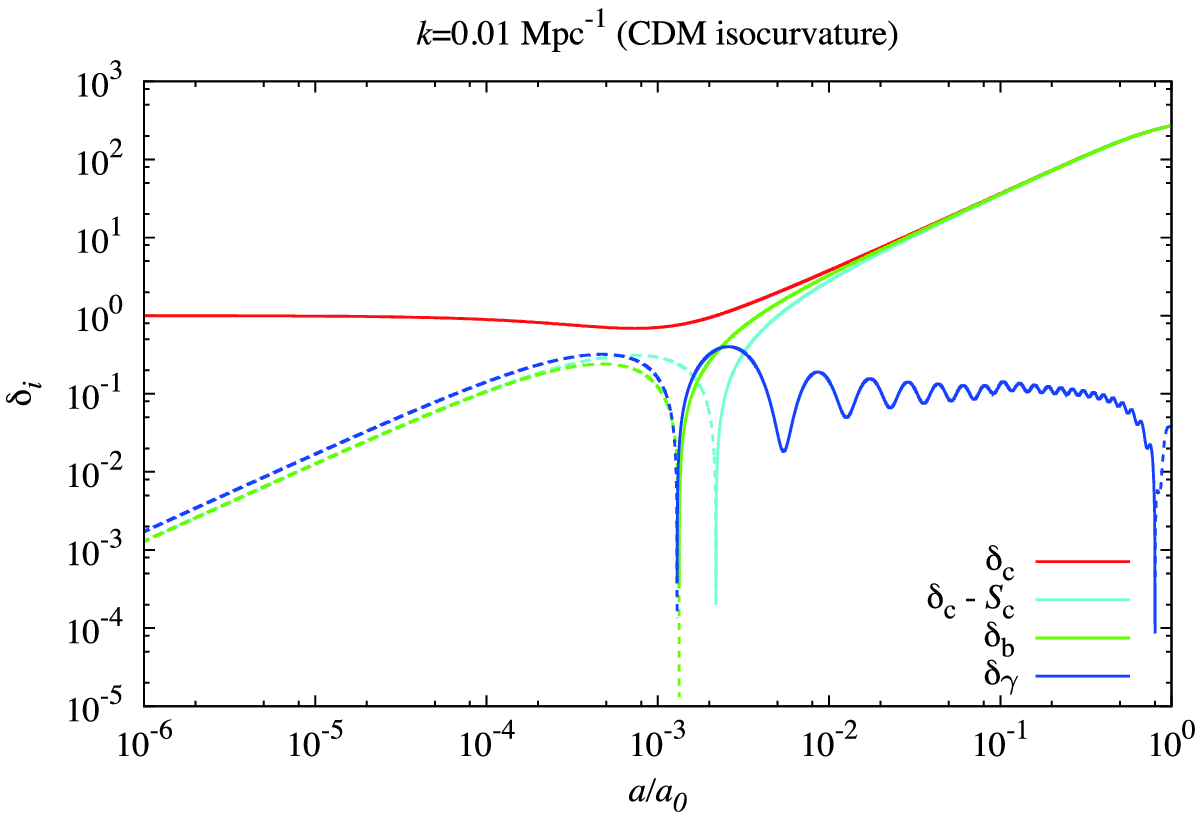}} &
      \hspace{0mm}
      \resizebox{80mm}{!}{\includegraphics{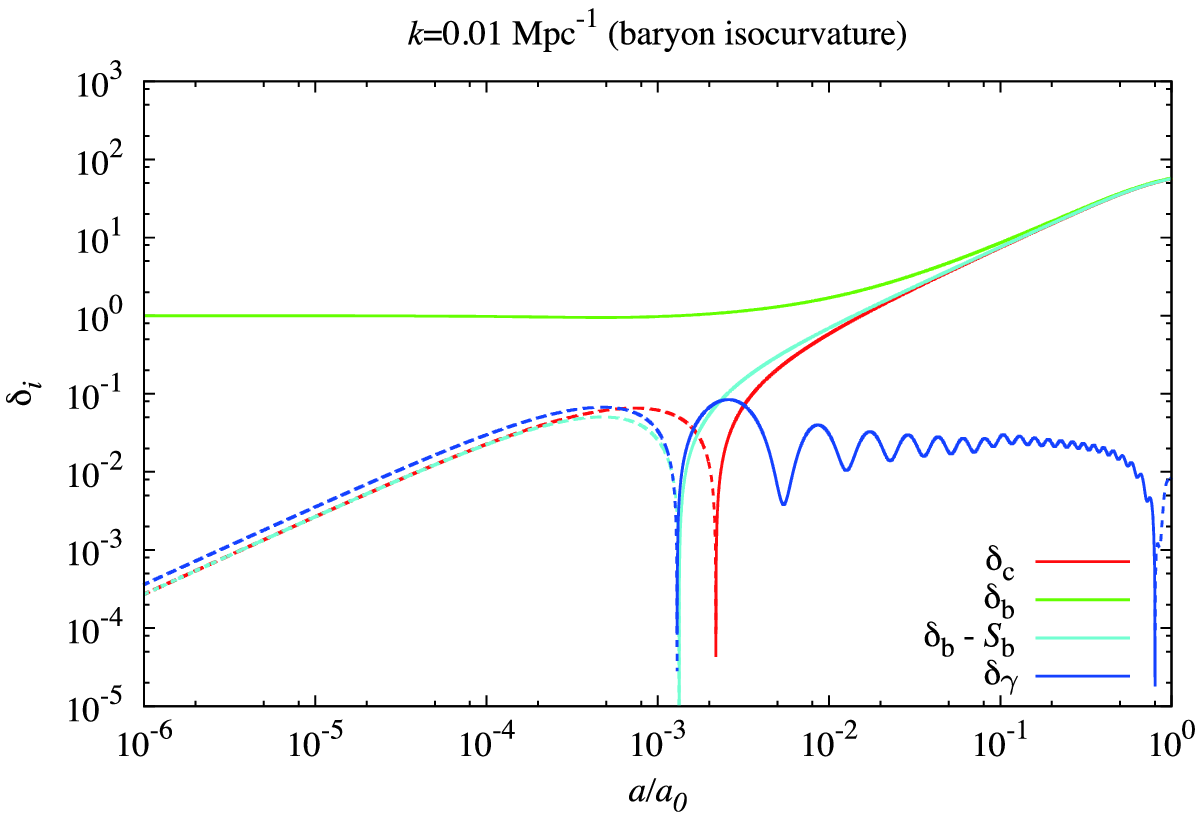}}  
    \\
      \resizebox{80mm}{!}{\includegraphics{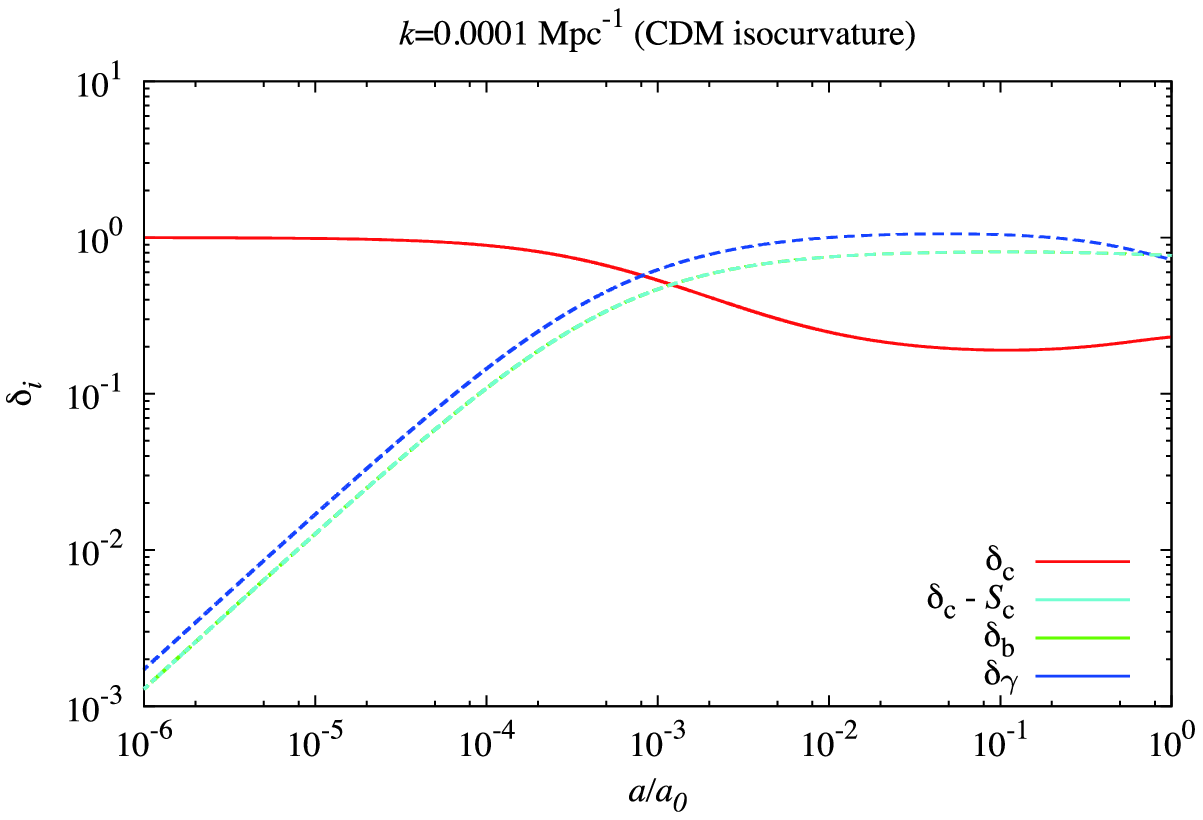}} &
      \hspace{0mm}
      \resizebox{80mm}{!}{\includegraphics{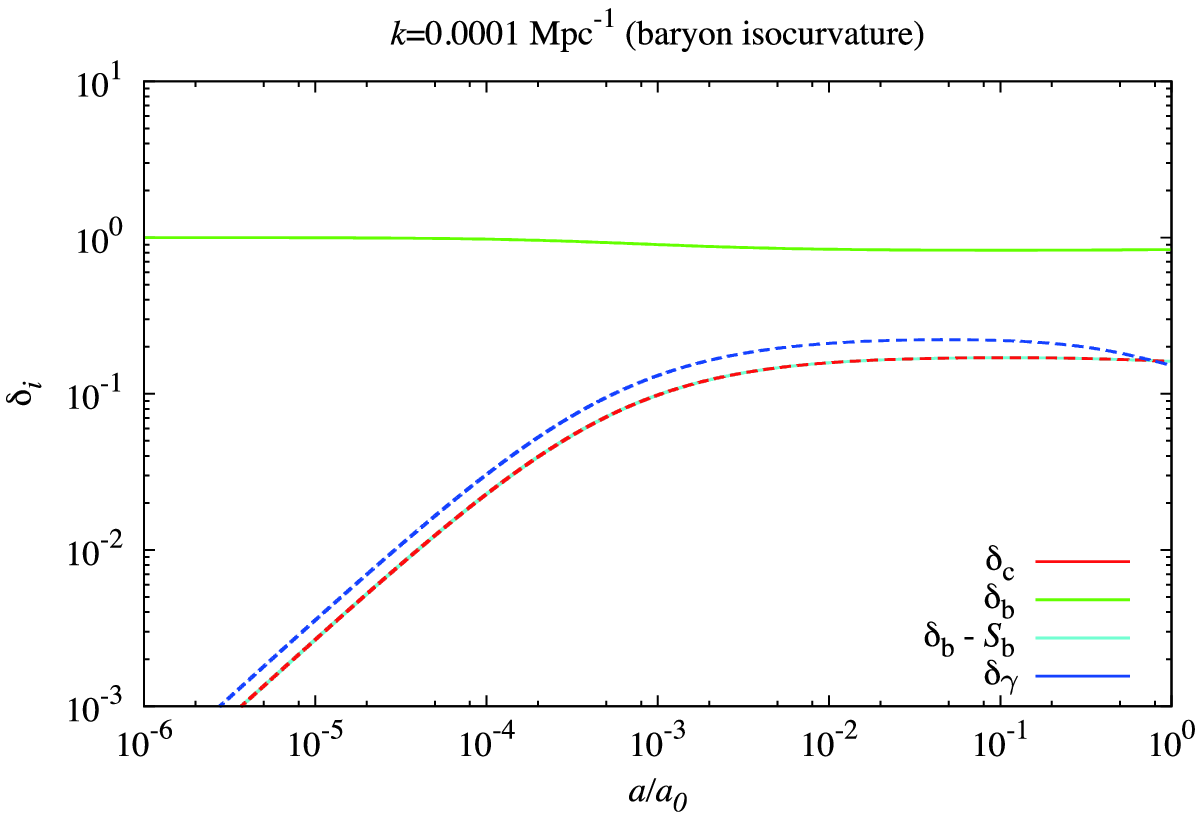}}      
    \end{tabular}
  \end{center}
  \caption{Evolutions of $\delta_c, \delta_b$ and  $\delta_\gamma$ for CDM isocurvature mode (left panel) 
  and baryon one (right panel)   in the synchronous gauge.  Evolutions of $\delta_c - S_c$ 
  and $\delta_c - S_b$ for CDM and baryon isocurvature modes respectively are also plotted.
  Here we show those for $k=1.0$ (top), $0.01$ (middle)
  and $0.0001~\mathrm{Mpc}^{-1}$ (bottom).}
  \label{fig:evolv}
\end{figure}

Now we discuss the evolutions of density fluctuations for each mode by showing those of 
CDM, baryon and photons. 
We start with the adiabatic mode. 
To identify the adiabatic mode,  it is convenient to introduce the gravitational (Newtonian) 
potentials $\phi$ and $\psi$ which are defined as the metric perturbations 
in the conformal Newtonian gauge:
\begin{equation}
ds^2 
= 
a^2 (\tau) \left[  -d \tau^2 (1 + 2\phi) + ( 1- 2 \psi)  \delta_{ij} dx^i dx^j \right].
\end{equation}
The metric perturbations $h$ and $\eta$ in the synchronous gauge 
can be related to $\phi$ and $\psi$ as 
\begin{equation}
\phi = 
\frac{1}{2k^2} \left[  
\ddot{h} + 6 \ddot{\eta} + \mathcal{H}  (\dot{h} + 6 \dot{\eta} )
\right],~~~~
\psi =  \eta - \frac{\mathcal{H}}{2k^2}  (\dot{h} + 6 \dot{\eta} ).
\end{equation}
In the previous section, we have defined the primordial power spectrum 
for the adiabatic mode with the curvature perturbation $\zeta$, which is 
related to $\psi$ and $\phi$ as 
\begin{equation}
\zeta = \psi+\frac{2}{3(1+w)}\left[\phi+\frac{\dot\psi}{\mathcal H}\right],
\end{equation}
where $w=\bar P/\bar\rho$ is the equation of state for the total component.

The adiabatic mode can be identified with a non-zero 
value of $\zeta$ for $\tau \rightarrow 0$, which leads to a solution on
superhorizon scales as, at leading order 
in the series of $k\tau$ \cite{Bucher:1999re}, 
\begin{eqnarray}
h/\zeta(0) & = &  \frac{1}{2} k^2 \tau^2, \\
\eta / \zeta (0) & = &  1  - \frac{5 + 4 f_\nu}{12(15+4f_\nu)} k^2 \tau^2, \\
\delta_c = \delta_b =  \frac{3}{4} \delta_\gamma =  \frac{3}{4} \delta_\nu
&=& - \frac{1}{4} k^2 \tau^2  \zeta (0), \\
\theta_\gamma=\theta_b&=&-\frac{1}{36}k^4\tau^3\zeta(0), \\
\theta_\nu/\zeta(0)&=&-\frac{1}{36}
\left[\frac{4f_\nu+23}{4f_\nu+15}\right]k^4\tau^3,
\end{eqnarray}
with the fraction of neutrinos in the energy density of radiations being denoted as
$f_\nu = \frac{\Omega_\nu}{\Omega_\gamma + \Omega_\nu}$.
Here the conformal time $\tau$ is normalized such that the scale factor is given 
by $a(\tau) = \tau + \tau^2$ 
with the normalization for $a$ being $a_{\rm eq} =1/4$. 
Although our main focus in this section is the evolutions of density fluctuations in 
models with isocurvature modes, for reference, we show those in the adiabatic mode 
in Fig.~\ref{fig:evolv_adi}. In the figure, cosmological parameters are taken as 
$\Omega_bh^2 (\equiv  \omega_b) = 0.0227, 
\Omega_ch^2 (\equiv \omega_c) = 0.108$, $h= 0.724$ and $\tau_{\rm reion}= 0.089$, 
where $\Omega_i = \rho_i(a_0) / \rho_{\rm crit}$ is density parameter for a species $i$, 
$h$ is the Hubble parameter and $\tau_{\rm reion}$ is the optical depth for reionization.
Unless otherwise stated, we assume these values in this paper.

Next we look at the evolutions of density perturbations in the CDM  isocurvature mode. 
This mode  can be identified with a non-zero value of $\delta_c$ at $\tau \rightarrow 0$, which we denote 
as $S_c$.  Notice that even in this mode, the evolutions of $\delta_i$ and $h$
follow Eq.~\eqref{eq:delta_diff}. Then the superhorizon solution in early times can be 
given by \cite{Bucher:1999re}
\begin{eqnarray}
h/S_c & = &  4 \Omega_c \tau - 6 \Omega_c \tau^2, \label{eq:CI_h}\\
\eta /S_c & = &  -\frac{2}{3} \Omega_c \tau + \Omega_c \tau^2, \\
\delta_c /S_c &=& 1 -2 \Omega_c \tau + 3 \Omega_c \tau^2,\\
\delta_b =  \frac{3}{4} \delta_\gamma =  \frac{3}{4} \delta_\nu
&=&  (-2 \Omega_c \tau + 3 \Omega_c \tau^2) S_c,\\
\theta_\gamma=\theta_b=\theta_\nu&=&-\frac{1}{3}\Omega_ck^2\tau^2S_c.
\label{eq:CI_theta} 
\end{eqnarray}
Notice that the metric perturbations $h$ and $\eta$ are induced by the existence of 
the initial isocurvature fluctuations. Importantly, once the metric perturbations are induced, 
density fluctuations are also generated through gravity, which satisfy the adiabatic condition
as for those of the adiabatic mode. Thus the superhorizon solution of $\delta_c$ for the CDM 
isocurvature mode has the structure of the form:
\begin{equation}
\delta_c  = (\textrm{initial fluctuations}) + (\textrm{induced from metric perturbations}).
\end{equation}
In the left panels of Fig.~\ref{fig:evolv}, we show the 
evolutions of $\delta_c,~\delta_b,~\delta_\gamma$ as well as 
$\delta_c - S_c$ which corresponds to the part induced by the metric perturbations in this mode.

On the other hand, the baryon isocurvature mode can be identified 
as a mode where $\delta_b$ does not vanish in the early times. The superhorizon 
solution in this mode is
\begin{eqnarray}
h /S_b & = &  4 \Omega_b \tau - 6 \Omega_b \tau^2, \label{eq:BI_h}\\
\eta /S_b& = &  -\frac{2}{3} \Omega_b \tau + \Omega_b \tau^2, \\
\delta_b /S_b&=& 1-2 \Omega_b \tau + 3 \Omega_b \tau^2,\\
\delta_c = \frac{3}{4} \delta_\gamma =  \frac{3}{4} \delta_\nu
&=&  (-2 \Omega_b \tau + 3 \Omega_b \tau^2 )S_b, \\
\theta_\gamma=\theta_b=\theta_\nu&=&-\frac{1}{3}\Omega_bk^2\tau^2S_b,
\label{eq:BI_theta} 
\end{eqnarray}
where the initial value of $\delta_b$ is denoted as $S_b$.
The superhorizon solution for baryon isocurvature mode above can be obtained by 
just replacing $\Omega_c$ in those for the CDM one with $\Omega_b$ (and $S_c$ with $S_b$)
except that $\delta_b$ and $\delta_c$ are replaced with each other.
Thus as far as $\delta_\gamma$ is concerned, its evolutions in CDM and baryon 
isocurvature modes are the same besides some
overall factors such as $\Omega_c$ and $\Omega_b$
originating from the amount of CDM and baryon (see Fig.~\ref{fig:evolv}).

So far we have seen that the initial, superhorizon solutions
of all perturbed quantities other than $\delta_c$ and $\delta_b$
differ only by an overall constant between the CDM and baryon
isocurvature modes. 
In fact, this is also true for all subsequent times, 
which was also noted in \cite{Gordon:2002gv} previously.
Here we discuss this issue in some details.

First of all, the superhorizon (initial) solutions of all perturbation quantities other than 
$\delta_c$ and $\delta_b$ are determined by
 the initial total matter isocurvature perturbation
$S_m=f_cS_c+f_bS_b$, not by $S_c$ and $S_b$ separately 
(here $f_c=\Omega_c/\Omega_m$, $f_b=\Omega_b/\Omega_m$ and 
 $\Omega_m=\Omega_c+\Omega_b$), which can be noticed 
by rewriting the superhorizon solutions with $S_m$: 
\begin{eqnarray}
h&=&\left[4\Omega_m\tau-6\Omega_m\tau^2\right]S_m,\\
\eta&=&\left[-\frac{2}{3}\Omega_m\tau+\Omega_m\tau^2\right]S_m,\\
\delta_\gamma=\delta_\nu&=&\left[-\frac{8}{3}\Omega_m\tau+4\Omega_m\tau^2\right]S_m,\\
\theta_\gamma=\theta_b=\theta_\nu&=&-\frac{1}{3}\Omega_mk^2\tau^2S_m,\\
\delta_m&=&\left[1-2\Omega_m\tau+3\Omega_m\tau^2\right]S_m,
\end{eqnarray}
where $\delta_m=f_c\delta_c+f_b\delta_b$.
(Pure) CDM and baryon isocurvature modes correspond to 
the cases of $S_b=0$ and $S_c=0$, respectively.
Secondly, the time derivatives of the perturbed quantities of any fluids
are dependent on $\delta_c$ and 
$\delta_b$ only through the metric perturbations $\dot h$ and hence the
matter perturbation $\delta_m$ (See Eqs.~(\ref{eq:deltac}-\ref{eq:deltanu})). 
Therefore, as far as the value of $S_m$ is fixed,  regardless of the values of $S_c$ and $S_b$, 
the subsequent time evolutions of all perturbed quantities apart from $\delta_b$ and $\delta_c$ 
are the same. For instance, as can be seen (except for the overall normalization) 
in Fig.~\ref{fig:evolv}, the evolution of $\delta_\gamma$
is the same for the CDM and baryon isocurvature modes.
This explains that models with the same $S_m$  give
identical CMB power spectra.
Therefore, in the linear cosmological perturbation theory, 
the CDM and baryon isocurvature modes are in principle
indistinguishable from observations of CMB. 
Furthermore, they also cannot be distinguished by observations of neither 
the metric nor the total matter perturbations.

Put in another way, the above  can also be understood as follows.
From a naive expectation, one may ask
why $\delta_c$ and $\delta_b$ would affect the perturbation evolutions 
in the same way even though baryon couples to photons while CDM does not. 
The reason is twofold. 
First, photons and baryon couple via the elastic Thomson scattering
and the momentum is just transferred between photons and baryons.
Therefore the photon perturbations are affected by $\theta_b$, 
but not by $\delta_b$ (See Eq.~\eqref{eq:theta_g}).
Second, both CDM and baryon are pressureless matter, thus the
velocity perturbations of these fluids $\theta_c$ and $\theta_b$ are
not generated from the density perturbations $\delta_c$ and $\delta_b$
(See Eq.~\eqref{eq:theta_b}),
which is not true for photons or neutrinos
(See Eqs.~\eqref{eq:theta_g} and \eqref{eq:deltanu}). 
These two facts explain that $\delta_b$
and $\delta_c$ affect the evolution of cosmological perturbations only 
through gravity in the same manner.

While the CDM and baryon isocurvature modes 
cannot be distinguished from the perturbations in the photon
fluid nor the metric, 
the evolutions of $\delta_c$ and $\delta_b$ are different between 
these two modes as seen from Fig.~\ref{fig:evolv}. 
Although the evolutions of the induced parts of $\delta_c$ and $\delta_b$  (represented 
as $\delta_c - S_c$ and $\delta_b - S_b$ in Fig.~\ref{fig:evolv}) from
 the metric perturbations are also the same except some constant factor 
 and retain the relation implied by Eq.~\eqref{eq:delta_diff},  
  the initial nonzero fluctuations of $\delta_c$ and $\delta_b$ for CDM and baryon isocurvature modes
  should be superimposed, respectively, on those induced parts. 
Thus 
 the evolutions of $\delta_c$ and $\delta_b$ are different, which gives a possibility of 
differentiating these modes. Although the difference between $\delta_c$ and $\delta_b$ 
becomes small on small scales since baryon density fluctuations catch up those of dark matter 
after the recombination and entering the horizon. 
However, on large scales, as seen from the bottom panels of Fig.~\ref{fig:evolv}, 
we can see some differences between $\delta_c$ and $\delta_b$ even at later epoch. 
Interestingly, we can in principle observe this difference 
by looking at 21 cm fluctuations which effectively probe the fluctuations of baryon. 
In the next section, we discuss this issue in some details.

\section{CDM and baryon isocurvature fluctuations with 21 cm Survey}

\subsection{21 cm fluctuations}
First we briefly summarize the fluctuations in redshifted 21 cm line emission.
Even after recombination at $z\simeq1090$, the photon and the hydrogen gas 
are still in kinetic equilibrium via the Compton scattering 
due to a small fraction of free electrons.
At around a redshift $z\simeq300$, the interaction becomes ineffective
and thereafter the gas cools adiabatically. 
The gas temperature $T_\mathrm{gas}$ scales as $a^{-2}$, 
while the photon temperature $T_\gamma$ scales as $a^{-1}$.
At first, the 21 cm spin flip driven by the atomic collision is so 
frequent that the spin temperature $T_s$\footnote{
$T_s$ determines the 
number ratio of neutral hydrogens in the triplet and singlet states, 
$n_\mathrm{triplet}/n_\mathrm{singlet}=3e^{-E_*/T_s}$,
where $E_*$ is the energy difference between the singlet and
triplet states.
} tracks $T_\mathrm{gas}$.
As gas cools faster than CMB, 
the atomic collision subsequently becomes less effective and
the interaction with the abundant CMB photons in turn dominates.
Then $T_s$ catches up with $T_\gamma$ until the formation 
of first objects, which is supposed to take place at $z\lesssim 30$.
Since $T_s$ remains lower than $T_\gamma$ throughout this
`dark age' at redshifts $30\lesssim z\lesssim 300$, 
the redshifted 21 cm line intensity is observed as an absorption 
compared with the CMB intensity at the microwave band
\cite{Scott:1990}.

The unperturbed brightness temperature of 21 cm line
observed now at an observed frequency $\nu$ 
can be calculated by \cite{Scott:1990,Madau:1996cs}
\begin{equation}
\bar T_b(\nu)=\left[
(1-e^{-\tau_\mathrm{21cm}})\frac{\bar T_s-\bar T_\gamma}{1+z}
\right](z_\nu)
\label{eq:barTb}
\end{equation}
where $z_\nu$ is the redshift of the 21 cm transition 
at the frequency $\nu$, i.e. $(1+z_\nu)= \nu_0/\nu$, 
with the frequency of 21 cm emission being denoted by
$\nu_0\simeq 1.4$ GHz.
$\tau_\mathrm{21cm}(z)$ is the optical depth of 21 cm, 
which can be given by
\begin{eqnarray}
\tau_\mathrm{21cm}(z)&=&
\left[\frac{3A_{10}\bar n_{\rm HI}
}{16 \nu_0^2\bar T_s (1+z) \mathcal H}\right](z)\notag\\
&=&1.081\times 10^{-2}
~\bar x_H\left[\frac{\bar T_s}{\bar T_\gamma}(z)\right]^{-1}
\left[\frac{\omega_b}{0.023}\right]
\left[\frac{\omega_m}{0.13}\right]^{-1/2}
\left[\frac{1+z}{10}\right]^{1/2},
\end{eqnarray}
where $A_{10}=2.85\times 10^{-15}$ s$^{-1}$ is the 
spontaneous emission coefficient, $\bar n_{\rm HI}$ 
is the number density of neutral hydrogen,
$x_{\rm HI}$ is the neutral hydrogen fraction, 
and $\omega_m\equiv \omega_b+\omega_c$.
In this paper we consider observations at
redshifts $z_\nu\gtrsim 30$ so that
complexities arising from nonlinear physics
can be much reduced.
In particular, we consistently set $\bar x_H=1$ at
these redshifts, adopting the 
standard assumption that the 
onset of the reionization occurs at latter epoch.
Then $\bar T_b(\nu)$ in Eq.~\eqref{eq:barTb} can be given by
\begin{equation}
\bar T_b(\nu)
=29.50\mbox{mK}
~\left[\frac{\bar T_s-\bar T_\gamma}{\bar T_s}(z_\nu)\right]
\left[\frac{\omega_b}{0.023}\right]
\left[\frac{\omega_m}{0.13}\right]^{-1/2}
\left[\frac{1+z_\nu}{10}\right]^{1/2}.
\end{equation}

Now we turn our attention to the fluctuations in $T_b$.
While there can be a number of sources generating 
the 21 cm brightness temperature fluctuations, 
the most dominant ones are as follows: (1)
the monopole source arising 
from the density fluctuations of neutral hydrogen
and relative temperature difference between 
the spin and CMB,  
(2) the Doppler redshift due to the peculiar velocity
of neutral hydrogen gas against the observer.
Taking into account these two effects, 
fluctuations in $T_b$ at frequency $\nu$ observed at position $\vec r_0$ 
and conformal time $\tau_0$ in a sight direction $\hat n$ 
can be given by \cite{Lewis:2007kz}
\begin{equation}
\Delta T_b(\nu,\hat n)=
\bar T_b(\nu)e^{-\tau_\mathrm{reion}}
\left[
\Delta_{n_\mathrm{HI}}+\frac{\bar T_\gamma}{\bar T_s-\bar T_\gamma}
(\Delta_{T_s}-\Delta_{T_\gamma})
+\frac{\tau_\mathrm{21cm}}{e^{\tau_\mathrm{21cm}}-1}
\hat n\cdot \frac{d\vec v/d\tau}{\mathcal H}
\right] (\vec r_\nu,\tau_\nu),
\label{eq:deltaTb}
\end{equation}
where $\Delta_X\equiv (X-\bar X)/\bar X$ represents 
the fractional perturbation of a quantity $X$ in the Newtonian gauge.
Eq.~\eqref{eq:deltaTb} is formal solution from the line of sight integration, 
and the right hand side should be evaluated at position 
$\vec r_\nu=\vec r_0+\hat n(\tau_0-\tau_\nu)$ and time $\tau=\tau_\nu$, 
where $\tau_\nu$ is the conformal time at $z_\nu$.
As shown in \cite{Lewis:2007kz}, in a redshift range $z\gtrsim 30$, 
other sources of 21 cm brightness temperature fluctuations 
are not very significant (at most percent-level at very large angular scales), 
as long as narrow redshift window functions are adopted.
Therefore in the following, we adopt Eq.~\eqref{eq:deltaTb}
as the theoretical prediction of observable 21 cm fluctuations.

The observed fluctuations in 21 cm intensity 
are determined by the fluctuations in the spin temperature $\Delta T_s$,
as well as $\Delta_{n_\mathrm{HI}}$, $\Delta_{T_\gamma}$ and $\vec v$.
Since the rates of the spin flip transition due to atom collisions and photon interaction
are sufficiently high, the evolution of $\Delta T_s$ 
can well be determined by assuming equilibrium.
For a comprehensive discussion for the evolution equations for 
$\Delta T_s$ and other fluctuations related for 21 cm fluctuations, 
we refer to \cite{Lewis:2007kz}.

\subsection{3-dim power spectrum from 21 cm tomography observations}

21 cm photons emitted at some
redshift $z$ now would be observed 
at a distinct  frequency $\nu=\nu_0/(1+z)$.
This allows us to perform
tomographic reconstruction of 
3-dim power spectrum of redshifted 21 cm fluctuations
\cite{Scott:1990, Kumar:1995, Madau:1996cs}.

Let us consider a tomographic observation of 
21 cm fluctuations around a fixed frequency $\nu$ and 
a direction $\hat n$ in the sky. 
The observed frequency and sky position constitute a 3-dim space
and we denote its coordinate as $(\delta\nu,\vec\Theta)$.
Once given a background cosmology, 
there is a one-to-one correspondence
between $(\delta\nu,\vec\Theta)$
and the coordinate of 3-dim real space 
$(r_\parallel,\vec r_\perp)$:
\begin{equation}
r_\parallel=y(z_\nu)\delta\nu,
\quad\vec r_\perp=d_A(z_\nu)\vec\Theta, 
\end{equation}
where the subscripts $\parallel$ and $\perp$ represent the
direction parallel and transverse to the line of sight, respectively.
The conversion factors between $(\delta \nu,\vec \Theta)$ and $(r_\parallel,r_\perp)$ 
are given by 
\begin{eqnarray}
y(z)&=&\frac{(1+z)^2}{\nu_0H(z)}, \\
d_A(z)&=&\int^z_0\frac{dz^\prime}{H(z^\prime)}. 
\end{eqnarray}
In particular, $d_A(z)$ is the comoving angular diameter distance.
Then a Fourier dual of $(\delta\nu,\vec\Theta)$, which we denote as
$(u_\parallel,\vec u_\perp)$, can also be related 
to the wave vector $(k_\parallel,\vec k_\perp)$ via 
\begin{equation}
k_\parallel=u_\parallel/y(z_\nu),
\quad\vec k_\perp=\vec u_\perp/d_A(z_\nu).
\end{equation}

From the formal solution Eq.~\eqref{eq:deltaTb}, 
we obtain the power spectrum of 21 cm fluctuations
observed at around frequency $\nu$ and sight direction $\hat n$ in $\vec k$ space
\begin{equation}
P_{\Delta T_b}(\vec k;\nu,\hat n)=\bar T_b^2e^{-2\tau_\mathrm{reion}}
\left[P_{00}(k,z_\nu)-2\mu^2P_{01}(k,z_\nu)
+\mu^4P_{11}(k,z_\nu)\right],
\end{equation}
where $\mu$ is the cosine of angle between $\hat n$ and $\vec k$, i.e.
$\mu=\hat n\cdot \vec k/k$.
Here $P_{00}$, $P_{11}$, $P_{01}$ are
the power spectra of auto-correlations of monopole
$\Delta_0\equiv \Delta_{n_\mathrm{HI}}+\frac{\bar T_\gamma}{\bar T_s-\bar T_\gamma}
(\Delta T_s-\Delta T_\gamma)$ and dipole moments 
$\Delta_1\equiv \frac{\tau_{\rm 21cm}}{e^{\tau_{\rm 21cm}}-1}\frac{kv}{\mathcal H}$
of the source and their cross-correlation, i.e.
\begin{eqnarray}
\langle\Delta_0(\vec k,z)\Delta_0(\vec k^\prime,z)\rangle
&=&P_{00}(k,z)(2\pi)^3\delta^{(3)}(\vec k+\vec k^\prime), \\
\langle\Delta_1(\vec k,z)\Delta_1(\vec k^\prime,z)\rangle
&=&P_{11}(k,z)(2\pi)^3\delta^{(3)}(\vec k+\vec k^\prime), \\
\langle\Delta_0(\vec k,z)\Delta_1(\vec k^\prime,z)\rangle
&=&P_{01}(k,z)(2\pi)^3\delta^{(3)}(\vec k+\vec k^\prime).
\end{eqnarray}
By performing a Fourier transformation
of the two-point correlation function
of the 21 cm fluctuations 
$\langle \Delta T_b(\vec r) \Delta T_b(\vec r^{\,\prime})\rangle$, 
the power spectrum defined in the $\vec u$ space, 
$P_{\Delta T_b}(\vec u)$, can be related to one in the wave 
vector space via
\begin{equation}
P_{\Delta T_b}(\vec u,\nu)=
\frac{1}{d_A^2(z_\nu)y(z_\nu)}
P_{\Delta T_b}(\vec k,z_\nu).
\end{equation}
Note that due to the Doppler effect, the power spectrum of 21 cm fluctuations
$P_{\Delta T_b}(\vec u)$ is not isotropic; it is instead a function of both 
$u_\parallel$ and $u_\perp$ or, equivalently,
both $k_\parallel$ and $k_\perp$.

\begin{figure}[htbp]
  \begin{center}
    \resizebox{160mm}{!}{
    \includegraphics{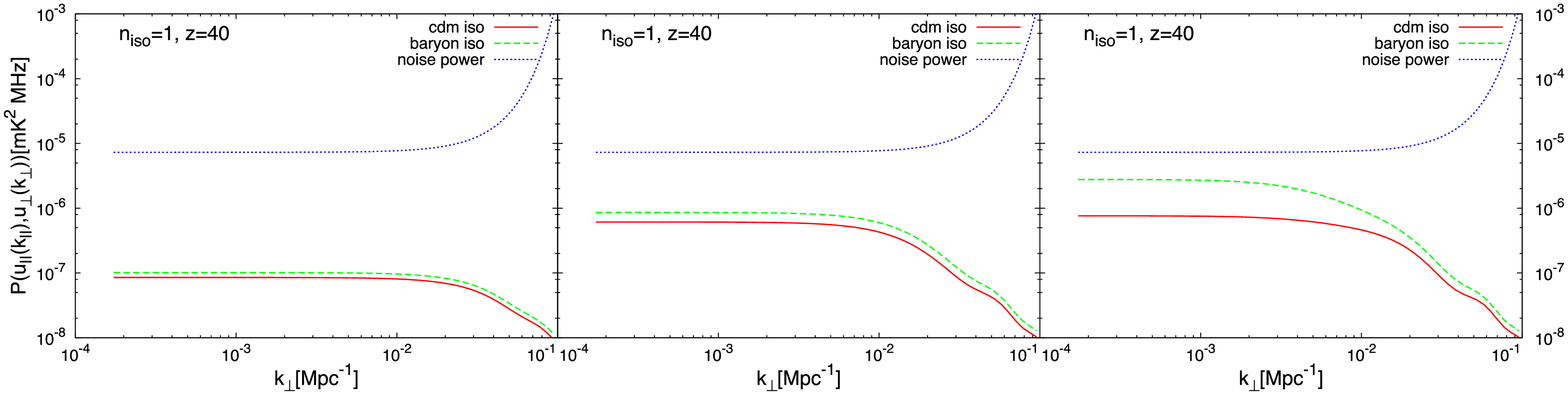} 
    }
    \resizebox{160mm}{!}{
    \includegraphics{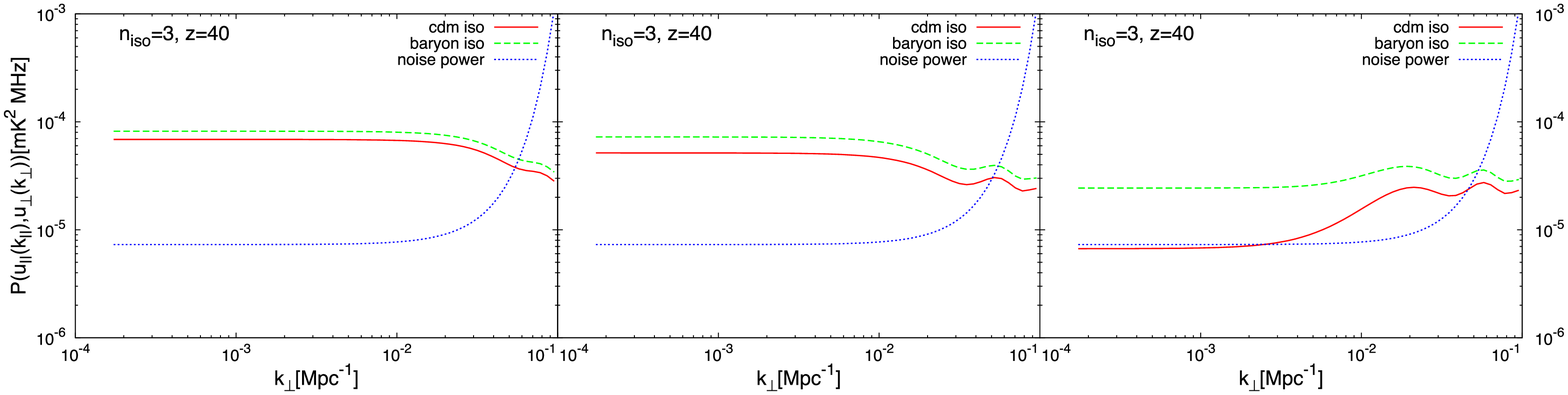} 
    }
  \end{center}
  \caption{21 cm power spectra $P_{\Delta T_b}(u_\parallel,u_\perp)$ 
  at $z=40$ for the cases with pure CDM (solid red line) and 
  baryon (dahsed green line) isocurvature modes. 
  The spectral indices are assumed as $n_s^{\rm (CDMiso)} = n_s^{\rm (biso)} =1$ (top) 
  and $n_s^{\rm (CDMiso)} = n_s^{\rm (biso)} =3$ (bottom). 
  In each panel, power spectra are shown as function of $k_\perp$ with
  $k_\parallel$ being fixed. From left to right, $u_\parallel$
  is fixed to 0.22MHz$^{-1}$, 0.72MHz$^{-1}$ and 2.3MHz$^{-1}$, 
  which respectively correspond to 
  $k_\parallel\simeq5.4\times 10^{-3}$Mpc$^{-1}$, 
  $1.7\times 10^{-2}$Mpc$^{-1}$ and $5.6\times 10^{-2}$Mpc$^{-1}$.
  The size of the baryon isocurvature mode is chosen such that 
  the baryon isocurvature mode gives the same CMB power spectrum with the CDM one.
  We also show the FFTT noise power spectra (dotted blue line).
  } 
  \label{fig:pk_21_1}
\end{figure}

In Fig.~\ref{fig:pk_21_1}, we show the 21 cm power spectra
for the cases with pure CDM and baryon isocurvature fluctuations.
To calculate the power spectra for 21 cm fluctuations, we used 
{\tt CAMB sources} \cite{Lewis:2007kz}. 
For reference, we also show expected sensitivity from the Fast Fourier Transform Telescope 
(FFTT) which can probe the redshift $z > 30$ and is described in \cite{Tegmark:2008au}.
An explicit form of noise power spectrum $P^{\rm noise}_{\Delta T_b}(\vec u)$ is given in 
Eq.~\eqref{eq:noiseP} and some detailed descriptions will be given in the next subsection.

Although any pure isocurvature modes are already excluded by current observations 
of CMB, we show pure isocurvature spectra to see to what extent we can 
tell the difference between CDM and baryon isocurvature modes.
In the top panels of Fig.~\ref{fig:pk_21_1}, we show power spectra for the case with $n_s^{\rm (CDMiso)} = n_s^{\rm (biso)}=1$.
The amplitude of primordial fluctuations for CDM mode is fixed as 
$\mathcal{P}_{S_c}(k_0) = 2.41 \times 10^{-9}$ at $k=0.002~{\rm Mpc}^{-1}$. 
For the baryon mode, the amplitude is assumed to give 
exactly the same CMB angular power spectrum as that for CDM 
(i.e., the initial amplitude is $(\Omega_c / \Omega_b)^2$ times large as that for CDM mode).
The bottom panels show the case of very blue-tilted spectra with 
$n_s^{\rm (CDMiso)} = n_s^{\rm (biso)}=3$. 
Although the spectral indices of this kind of value seem somewhat too large, it has been claimed that 
blue-tilted (CDM/baryon) isocurvature spectrum may be favored by observations 
\cite{Keskitalo:2006qv,Beltran:2004uv,Beltran:2005gr,Bean:2006qz,Sollom:2009vd,Li:2010yb}\footnote{
A recent analysis puts a constraint on the spectral index as
$n_s^{\rm (CDMiso)} = 2.246^{+ 0.494}_{- 0.428}$ \cite{Li:2010yb}.
}.
In addition, some axion model producing very blue-tilted spectrum has been discussed \cite{Kasuya:2009up},
and in their model, the spectral index for the CDM isocurvature perturbations
can be as large as 4, depending on parameters. 
In light of these, it would be interesting to investigate isocurvature models with very blue-tilted spectra. 

As seen from Fig.~\ref{fig:pk_21_1}, the survey of 21 cm fluctuations 
gives different power spectra for CDM and baryon isocurvature fluctuations, 
which enables us to discriminate these modes.
However, when the primordial spectra are assumed to be scale-invariant (top panels), 
the signals are too weak compared with sensitivities of observations.
On the other hand, 
for the case with $n_s^{\rm (CDMiso)} = n_s^{\rm (biso)}=3$ (bottom panels), 
the amplitudes become large and difference in power spectra can larger than 
the noise power. So we expect that these kinds of  isocurvature modes
can be differentiated with the 21 cm survey of FFTT.

\begin{figure}[htbp]
  \begin{center}
    \resizebox{160mm}{!}{
    \includegraphics{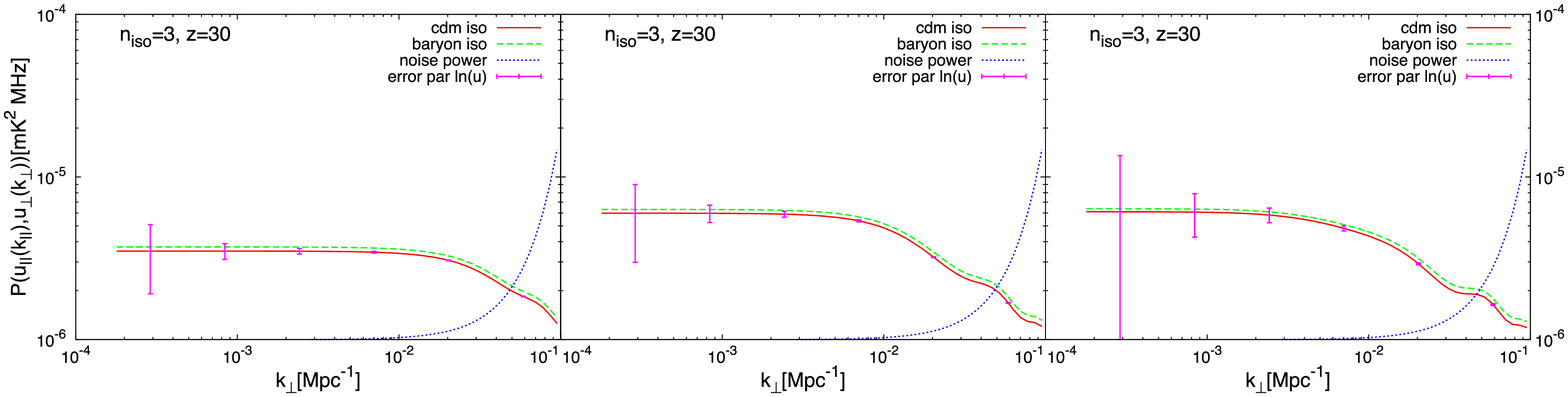} 
    }
    \resizebox{160mm}{!}{
    \includegraphics{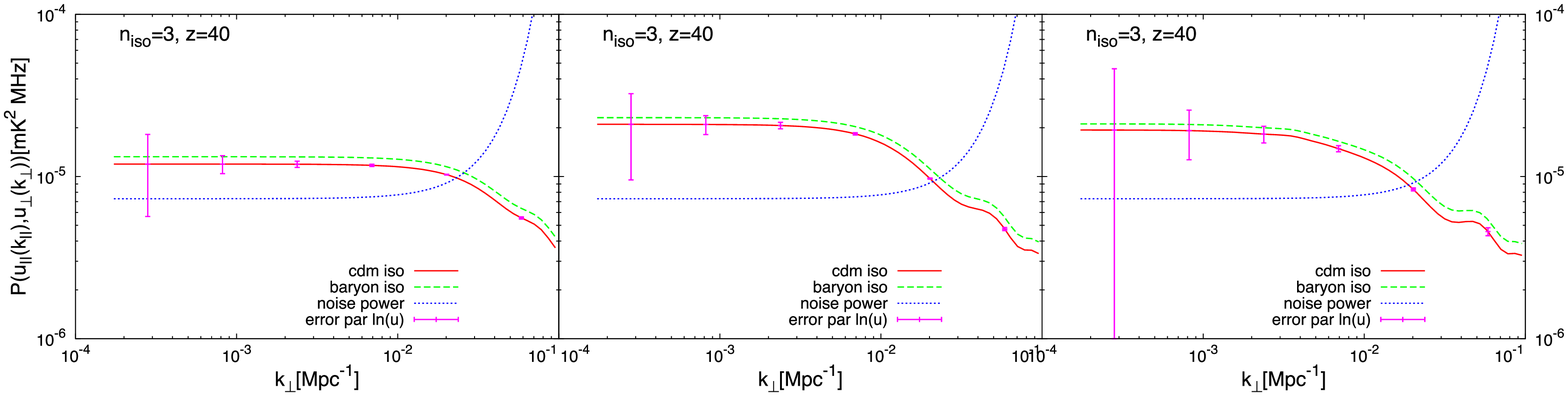} 
    }
    \resizebox{160mm}{!}{
    \includegraphics{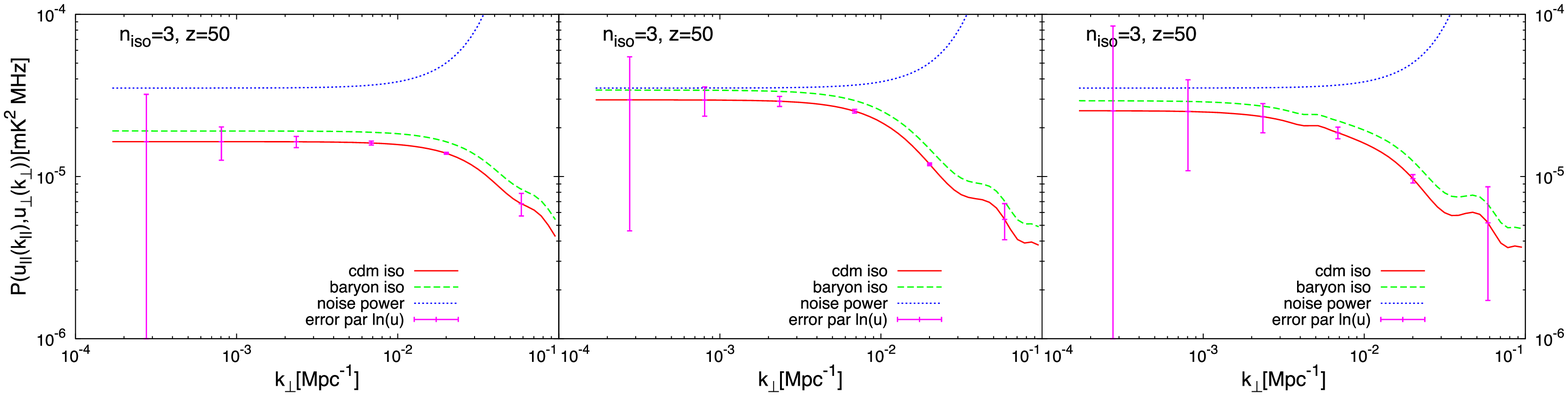} 
    }
  \end{center}
  \caption{
  21 cm power spectra $P_{\Delta T_b}(u_\parallel,u_\perp)$ 
  from the mixture of adiabatic and uncorrelated matter isocurvature perturbations.
  From top to bottom, we show power spectra at $z=30$, 40 and 50 in order.
  In each panel, power spectra are shown as a function of $k_\perp$, for a fixed $k_\parallel$.
  From left to right, $u_\parallel$
  is fixed to 0.22MHz$^{-1}$, 0.72MHz$^{-1}$ and 2.3MHz$^{-1}$, 
  which respectively correspond to 
  $k_\parallel\simeq5.4\times 10^{-3}$Mpc$^{-1}$, 
  $1.7\times 10^{-2}$Mpc$^{-1}$ and $5.6\times 10^{-2}$Mpc$^{-1}$.
  Shown are power spectra for the cases with $(r_c,r_b)=(0.1,0)$ (dashed green line)
  and $(r_c,r_b)=(0,0.1\times (\Omega_c/\Omega_b)^2)$ (dotted blue line).
  The spectral indices for the isocurvature
  fluctuations are assumed as $n_s^{\rm (CDMiso)} = n_s^{\rm (biso)} =3$.
  Notice that these two cases give completely the same CMB power spectrum.
  We also show the FFTT noise power spectra (dotted blue line) and 
  expected errors in measuring 21 cm power spectrum (magenta bar).
  Errors are estimated adopting a pixelization, which is logarithmically uniform in 
  $(u_\parallel, u_\perp)$ with widths $\Delta \ln u_\parallel\simeq\Delta \ln u_\perp\simeq 1$.
  } 
  \label{fig:pk_21_2}
\end{figure}

In fact, even when we consider a realistic case where CDM/baryon isocurvature fluctuations 
give subdominant contributions to the total (almost adiabatic) spectrum allowed by 
current observations, we can still probe the difference between CDM and baryon 
isocurvature modes. 
In Fig.~\ref{fig:pk_21_2}, we show the 21 cm spectra for $(r_c, r_b) = (0.1,0)$ (red line) and 
$(r_c, r_b) = (0, 0.1\times (\Omega_c/\Omega_b)^2)$ (green line). 
Contributions from the adiabatic perturbations are also included here.
The choice of the value of $r_b$ for the latter case is made so that 
these two cases give the indistinguishable CMB angular power spectra.
In the figure, we also show the expected error in measuring power spectrum
$\delta P_{\Delta T_b}(\vec u)$, as well
as that for the noise  $P^{\rm noise}_{\Delta T_b}(\vec u)$. 
An explicit expression of $\delta P_{\Delta T_b}(\vec u)$ is given in Eq.~\eqref{eq:deltaP}. 
While the noise power spectrum is estimated from the detector noise,  
error accounts for both the cosmic variance (assuming a vanilla model
with the WMAP 7yr mean parameters \cite{Komatsu:2010fb}) and survey setups. 
The figure indicates that these two cases can be differentiated 
with the FFTT survey. 
To see this issue more quantitatively, in particular, 
to what extent we can probe the difference with a future 21 cm tomography observation 
such as FFTT, we make a forecast by adopting the Fisher information matrix analysis 
in the next section. In closing this section, 
we also note that the the difference between 
the two isocurvature modes are more prominent 
at higher redshifts. This reflects the fact that 
the density perturbation in baryon catches up with 
one in CDM as redshift decreases, as we mentioned in the
previous section.

\subsection{Discriminating CDM and baryon isocurvature fluctuations with 
future 21 cm survey}

Now we make a quantitative discussion about the possibility of discriminating 
CDM and baryon isocurvature 
modes with future observations of 21 cm fluctuations.  
For this purpose, we assume 
FFTT \cite{Tegmark:2008au} for 21 cm survey.
Although 21 cm surveys can probe the difference between CDM and baryon isocurvature modes,
regarding other cosmological parameters, CMB observations are much more powerful. 
Thus we also include a future CMB observation, for which we assume 
the specification of CMBpol \cite{Baumann:2008aq} in the Fisher matrix analysis.

Before presenting our results, here we briefly describe the formalism of the Fisher 
matrix analysis. Since we use expected measurements of  CMB and 21 cm survey, 
the Fisher matrix $F_{ij}$ can be given by the sum of two contributions:
\begin{equation}
F_{ij} = F_{ij}^{\rm (CMB)} + F_{ij}^{\rm (21cm)},
\end{equation}
where indices $i,j$ represent cosmological parameters and $F_{ij}^{\rm (CMB)}$ 
and $F_{ij}^{\rm (21cm)}$ respectively indicate the contributions from CMB and 
21 cm observations. We analyze a forecast  in a 9 dimensional parameter space 
$(\omega_c, \omega_b, h, n_s, \tau_{\rm reion}, n_s, A_s,r_c, r_b, n_s^{\rm (CDMiso)})$. 
The definitions of the parameters here are already given in Section~\ref{subsec:evolution} 
except $A_s$, the amplitude of primordial fluctuations at the reference scale $k=0.002~{\rm Mpc}^{-1}$.
Here we assume that the spectral index for baryon isocurvature mode is the same 
as that for CDM one, i.e., $n_s^{\rm (biso)} = n_s^{\rm (CDMiso)}$ is adopted in the analysis.

\begin{table}
  \centering
  \begin{tabular}{|c|c|c|}
  \hline 
  Frequency (GHz) & beam  FWHM (arcmin) & $\Delta T$ ($\mu$K arcmin) \\
\hline
45   & 17 & 5.85  \\
70   & 11 & 2.96 \\
100 & 8   & 2.29 \\
150 & 5   & 2.21 \\
220 & 3.5 & 3.39 \\
\hline
\end{tabular} 
  \caption{Specifications for the mid-cost (EPIC-2m) CMBpol mission used 
  in the analysis. For polarization, $\Delta T$ is reduced by the factor of $\sqrt{2}$. 
  We have excluded the highest and lowest frequency bands (30 GHz and 340 GHz)
  for a realistic foreground removal.}
  \label{table:cmbpol}
\end{table}

For the Fisher matrix for CMB $F_{ij}^{\rm (CMB)}$, we follow the method of 
\cite{Zaldarriaga:1997ch} and assume  the projected CMBpol mission 
\cite{Baumann:2008aq}. More specifically, we use the specifications for 
the mid-cost (EPIC-2m) mission, which is given in Table~\ref{table:cmbpol}.

The Fisher matrix from a 21 cm
tomography survey can be given by \cite{Tegmark:1997rp,Mao:2008ug}, 
\begin{equation}
\label{eq:Fij21}
F_{ij}^{\rm (21cm)}=\int \frac{d^3u}{(2\pi)^3}
\frac{V_\Theta}{P^{\rm tot}_{\Delta T_b}(\vec u)^2}
\left(\frac{\partial P_{\Delta T_b}(\vec u)}{\partial \lambda_i}\right)
\left(\frac{\partial P_{\Delta T_b}(\vec u)}{\partial \lambda_j}\right),
\end{equation}
where $\lambda_i$ is the $i$-th cosmological parameter
and $P^{\rm tot}_{\Delta T_b}(\vec u)$
is the power spectrum of the total observed 21 cm fluctuations, 
which is the sum of the signal $P_{\Delta T_b}(\vec u)$
and the noise power spectra $P^{\rm noise}_{\Delta T_b}(\vec u)$, 
i.e. $P^{\rm tot}_{\Delta T_b}(\vec u)=P_{\Delta T_b}(\vec u)+
P^{\rm noise}_{\Delta T_b}(\vec u)$.
Here $V_\Theta=\Omega_{\rm FoV}B$ is the survey volume in 
$\Theta$ space, where $\Omega_{\rm FoV}$ and 
$B$ are the solid angle in the field of view and
the frequency band width, respectively.
In the analysis, we set $\Omega_{\rm FoV}=\pi$
and $B=8$\,MHz, regardless of observed frequencies.

Noise power spectrum of FFTT can be given by
\cite{Tegmark:2008au}
\begin{equation}
P^{\rm noise}_{\Delta T_b}(\vec u)=
\frac{4\pi f_{\rm sky} \lambda_\nu^2 T_{\rm sys}^2}{
A\Omega_{\rm FoV}f_{\rm cover}t_{\rm obs}}
W(\vec u_\perp)^{-2}.
\label{eq:noiseP}
\end{equation}
Here $f_{\rm sky}$ is the observed sky fraction, 
$\lambda_\nu$ is the observed radio wave length, 
$T_{\rm sys}$ 
is the system temperature, 
$A$ is the telescope area, 
and $f_{\rm cover}$ is the fraction of are covered by the array.
In the analysis, we take $f_{\rm cover}=f_{\rm sky}=1$,
$A=20\mbox{km}^2$ and $t_{\rm obs}=1\mbox{yr}$.
$W(\vec u_\perp)$ is the beam response function, which 
in principle depends on the shape of the telescope.
In the analysis, we assume a symmetric Gaussian beam
and we take
\begin{equation}
W(\vec u_\perp)=\exp\left[
-\frac{\lambda_\nu^2}{A}u_\perp^2
\right].
\end{equation}

Since both $P_{\Delta T_b}(\vec u)$ and $P^{\rm noise}_{\Delta T_b}(\vec u)$
do not depend on the azimuthal direction in $\vec u$-space, 
we integrate the right hand side in Eq.~\eqref{eq:Fij21}
over this direction.
In addition, we pixelize the $\vec u$ space
with widths $\Delta u_\parallel$ and 
$\Delta u_\perp$ in the line of sight 
and transverse directions, respectively.
Then Eq.~\eqref{eq:Fij21} can be approximated as 
\begin{equation}
\label{eq:Fij21app}
F_{ij}^{\rm (21cm)}\simeq
\sum_{\rm pixels}
\frac{1}{\delta P_{\Delta T_b}(\vec u)^2}
\left(\frac{\partial P_{\Delta T_b}(\vec u)}{\partial \lambda_i}\right)
\left(\frac{\partial P_{\Delta T_b}(\vec u)}{\partial \lambda_j}\right),
\end{equation}
where $\delta P_{\Delta T_b}(\vec u)$ is 
the error in power spectrum par pixel, which can be given by
\begin{equation}
\delta P_{\Delta T_b}(\vec u)
=\sqrt{\frac{(2\pi)^2}{
\Delta u_\parallel
u_\perp \Delta u_\perp V_\Theta}}
P_{\Delta T_b}^{\rm tot}(\vec u).
\label{eq:deltaP}
\end{equation}

Removal of foregrounds is one of the primary
challenges in future observation of cosmological 
21 cm fluctuations.
We assume foregrounds are highly smooth in the frequency space, 
and can be removed to sufficient accuracy at $|u_\parallel| \ge 1/B$, where
we assume $T_{\rm sys}=200$K$[(1+z)/10]^{2.6}$ is dominated by the sky
temperature due to the Galactic synchrotron emission.
Note that this choice of lowest $|u_\parallel|$ hardly affects our results, as 
we focus on a blue-tilted power spectra of isocurvature perturbations.
In addition, we also exclude modes $k>k_{\rm max}=0.1$ Mpc$^{-1}$
in the analysis, in order to neglect effects of the
nonlinear evolution of matter perturbations.

\begin{figure}[htbp]
  \begin{center}
  \begin{tabular}{cc}
    \hspace{-5mm}
    \resizebox{80mm}{!}{
    \includegraphics{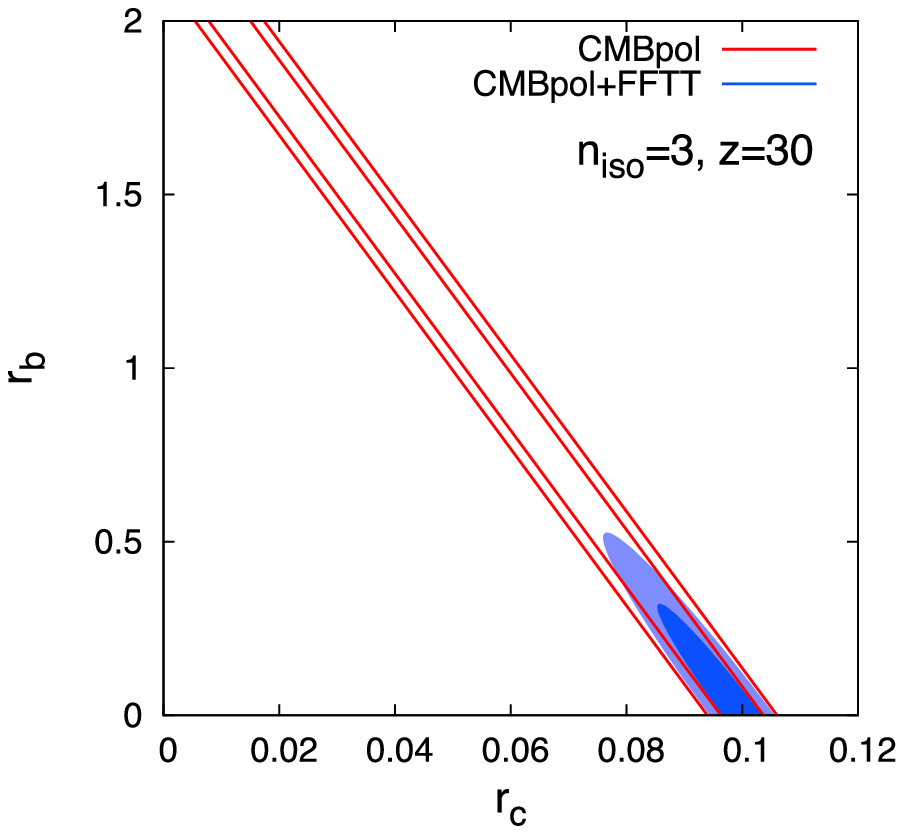}
    } & \hspace{-5mm}
    \resizebox{80mm}{!}{
    \includegraphics{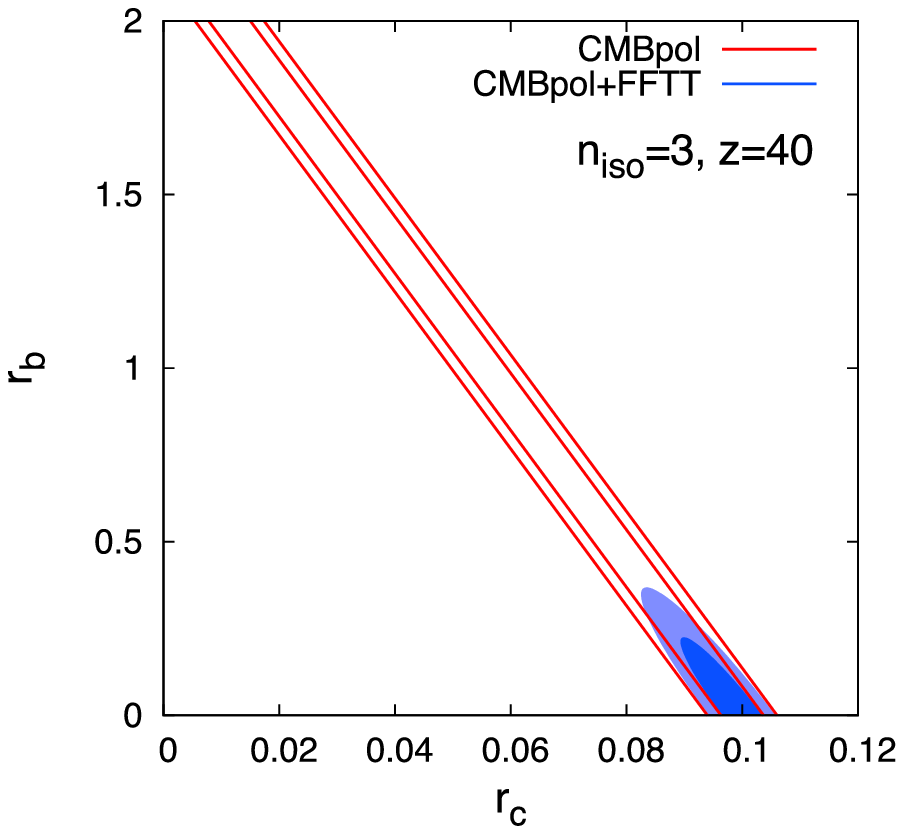}
    } \\
    \hspace{-5mm}
    \resizebox{80mm}{!}{
    \includegraphics{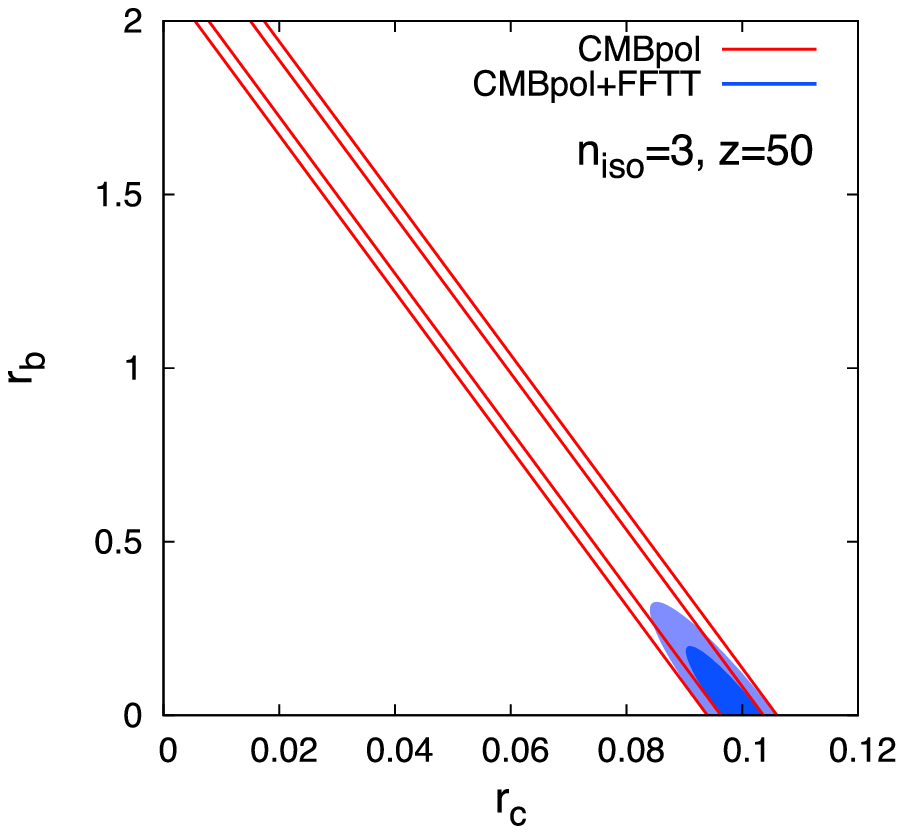}
    } & \hspace{-5mm}
    \resizebox{80mm}{!}{
    \includegraphics{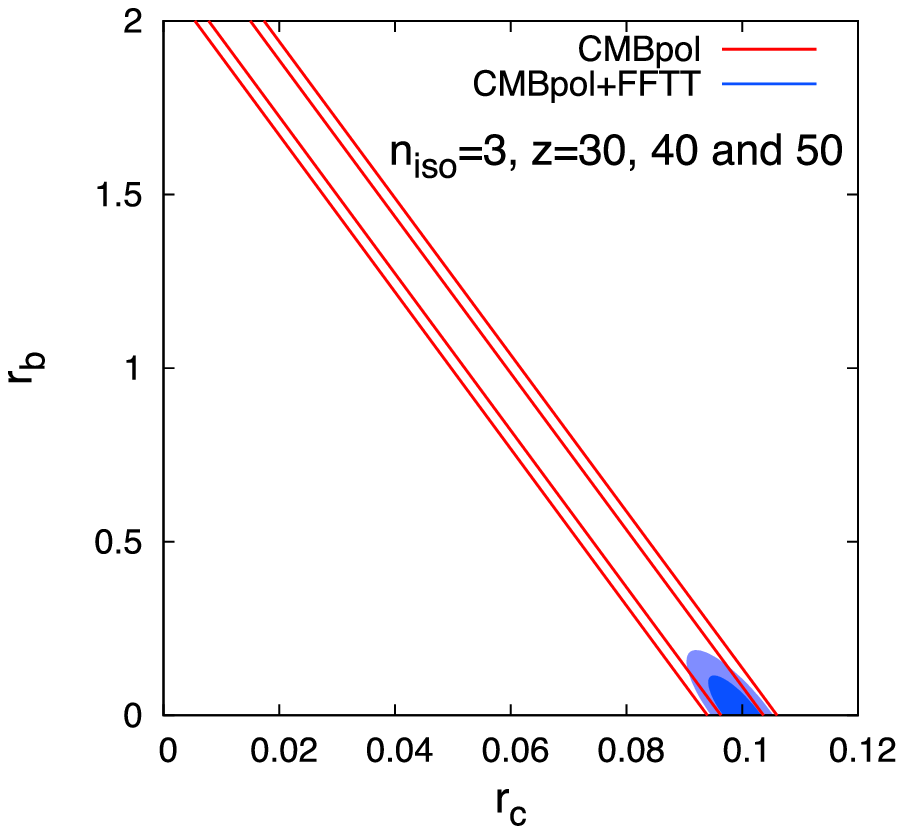}
    }
    \end{tabular}
  \end{center}
  \caption{Expected 1$\sigma$ and 2$\sigma$ constraints on $r_c$ and $r_b$ from CMBpol alone
  (red contours) and CMBpol+FFTT (shaded regions).
  As for FFTT, we show constraints from different combinations
  of observed redshifts in separate panels; 
  $z=30$ (top left), $z=40$ (top right), $z=50$ (bottom left)
  and all of these redshifts $z=30$, $40$ and $50$
  (bottom right).
  The fiducial values are assumed as $(r_c, r_b)=(0.1, 0)$ and 
  $n_s^{\rm (CDMiso)} = n_s^{\rm (biso)}=3$. 
  }
  \label{fig:pink_rc_rb}
\end{figure}

\begin{figure}[htbp]
  \begin{center}
  \begin{tabular}{cc}
    \hspace{-5mm}
    \resizebox{80mm}{!}{
    \includegraphics{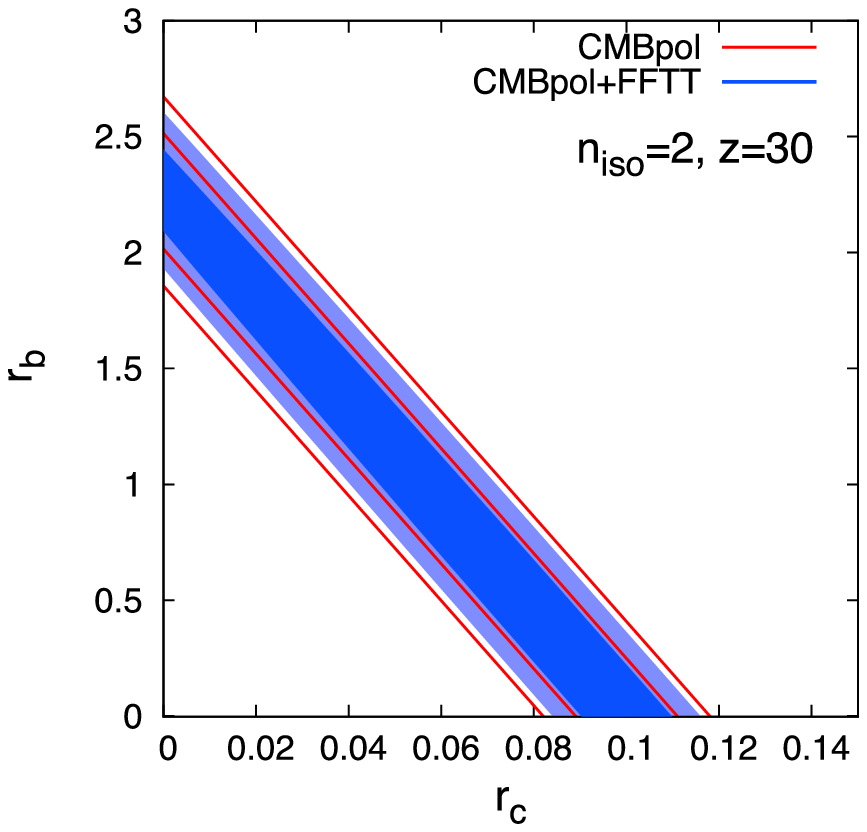}
    } & \hspace{-5mm}
    \resizebox{80mm}{!}{
    \includegraphics{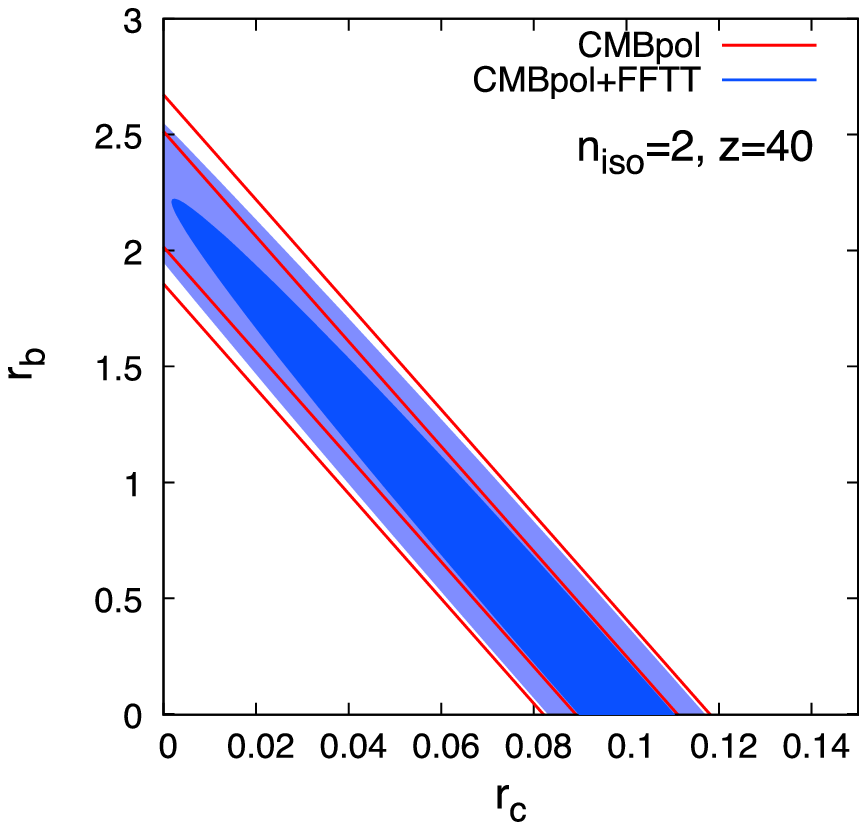}
    } \\
    \hspace{-5mm}
    \resizebox{80mm}{!}{
    \includegraphics{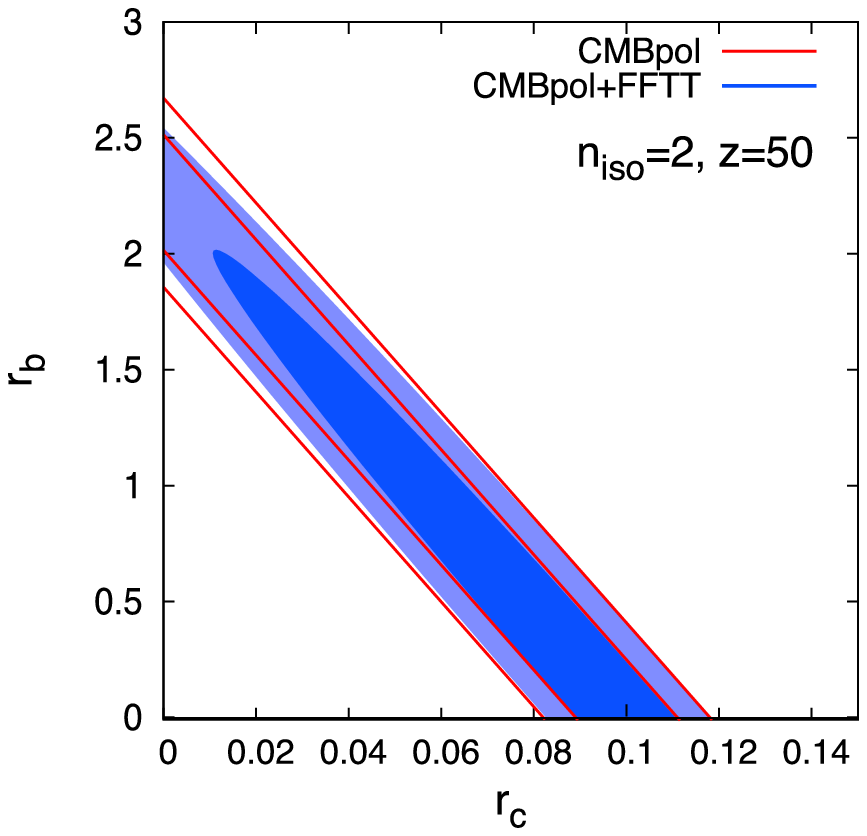}
    } & \hspace{-5mm}
    \resizebox{80mm}{!}{
    \includegraphics{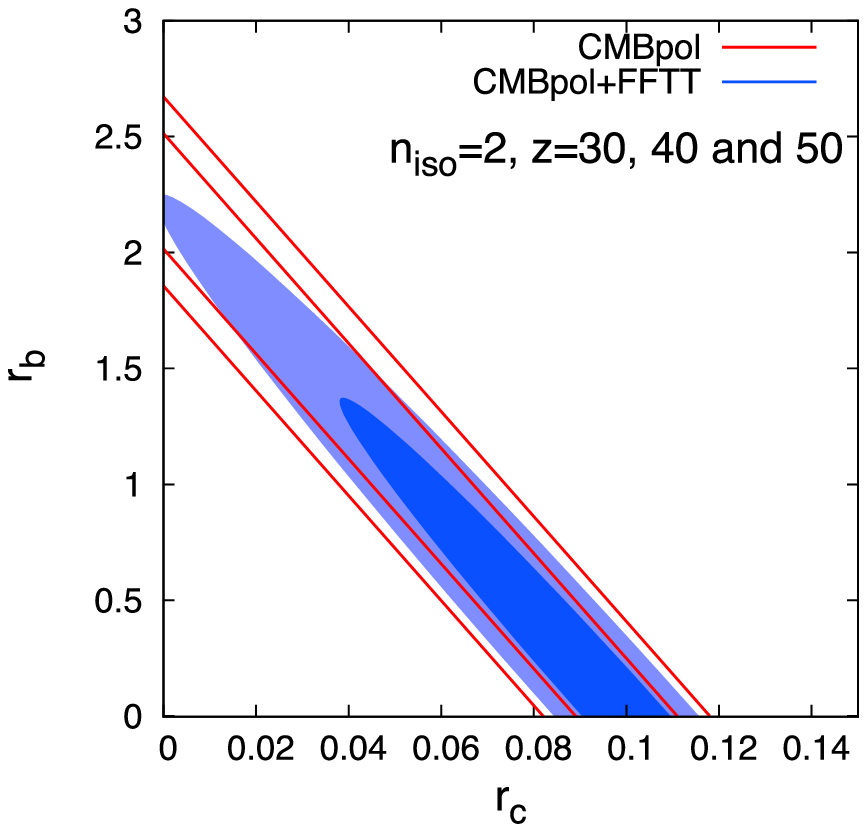}
    }
  \end{tabular}
  \end{center}
  \caption{Same as in Fig.~\ref{fig:pink_rc_rb} but for 
  $n_s^{\rm (CDMiso)} = n_s^{\rm (biso)}=2$. 
  }
  \label{fig:red_rc_rb}
\end{figure}

In Fig.~\ref{fig:pink_rc_rb}, we show projected constraints for $r_c$ and $r_b$
from CMBpol alone (solid red contours) and CMBpol+FFTT (shared regions).
Regarding the observed redshifts of FFTT survey, 
we analyzed four different combinations:
$z=30$ (top left), $z=40$ (top right)
$z=50$ (bottom left) and all these redshifts combined (bottom right).
Here the fiducial values are assumed as $r_c = 0.1$ and $r_b=0$ and 
$n_s^{\rm (CDMiso)} = n_s^{\rm (biso)}=3$. Other cosmological 
parameters  are assumed as those given in the previous section.
Since CMB alone cannot in principle make a 
distinction between CDM and baryon isocurvature models, the constraint 
from CMBpol alone is completely degenerate along the line which gives 
the same amount of isocurvature fluctuations in term of matter:
\begin{equation}
 \left( \frac{\Omega_c}{\Omega_m} \right)^2 r_c 
 + \left( \frac{\Omega_b}{\Omega_m} \right)^2 r_b = C,
\end{equation}
with $C = 0.1\times (\Omega_c/ \Omega_m)^2 $.
However, as one can see from the figure,  by including 21 cm surveys, the degeneracy 
can be removed, which means that we can see the difference between 
CDM and baryon isocurvature modes with 21 cm fluctuation surveys, in particular, 
when  isocurvature modes have very blue-tilted
spectra such as $n_s^{\rm (CDMiso)} = n_s^{\rm (biso)}=3$.

As discussed in the previous section, the difference between $\delta_c$ and $\delta_b$ 
is more prominent at higher redshifts. 
On the other hand, observational sensitivity is better for lower redshifts. 
For a fixed band width $B$ and a maximum wave number for 
the linear perturbation evolution $k_{\rm max}$, the increase in the signal overwhelms
the increase in noise and the most stringent constraint is obtained 
at the highest redshift $z=50$.

We also note that the constraint from the combination of three
redshifts $z=30$, $40$ and $50$ is at least twice as stringent as one from
a single redshift $z=50$. This suggest that a combination of 
different redshifts offers to some extent a synergy in discriminating 
the CDM and baryon isocurvature perturbations.
This comes from the fact that parameter degeneracies between
$r_c$ or $r_b$ and other cosmological parameters
more or less differ at different redshifts.
By combining observations at different redshifts, parameter degeneracies 
are partially removed and this gives severer constraints on
$r_c$ and $r_b$. Otherwise, the constraint from the combination
of three redshifts would be at most as $\sqrt 3$ times tighter as one from $z=50$, 
which is the most stringent if a single redshift is observed.

Finally we comment on the dependence of constraints 
on the spectral indices of primordial isocurvature power spectra.
In Fig.~\ref{fig:red_rc_rb}, we 
show constraints in the $r_c$-$r_b$ plane for the case of 
$n_s^{\rm (CDMiso)} = n_s^{\rm (biso)}=2$.
As can be expected from the discussion in the previous section, 
constraints on $r_c$ and $r_b$ significantly degrade
as the spectral indices becomes smaller. 
For the $n_s^{\rm (CDMiso)} = n_s^{\rm (biso)}=2$, 
we may marginally discriminate the CDM and baryon 
isocurvature perturbations for a fiducial model $(r_c,r_b)=(0.1, 0)$.
The significance is at around 1 $\sigma$ level from an observation 
at single redshift $z=50$. By combining observations at all
redshifts, the significance can be improved to 
around 2 $\sigma$ level.
However, note that to what extent we can discriminate 
the CDM and baryon isocurvature perturbations 
depends on the maximum wave number for linear evolution.
While we adopted a fixed $k_{\rm max}=0.1$\,Mpc$^{-1}$ at
any observed redshifts, the use of larger $k_{\rm max}$ may be allowed
as redshift becomes higher, which may
offer more stringent constraints on $r_c$ and $r_b$.
On the other hand, for the case of $n_s^{\rm (CDMiso)} = n_s^{\rm (biso)}=1$
we found FFTT does not improve constraints on $r_c$ and $r_c$ from CMBpol alone
regardless of observed redshifts. 
This is what we expect from the top panels in Fig.~\ref{fig:pk_21_1}.
Thus the CDM and baryon isocurvature perturbations 
would be hardly distinguished if their power spectra are nearly scale invariant.

\section{Summary}

In this paper, we discussed a possibility of differentiating 
CDM and baryon isocurvature fluctuations with 21 cm survey.  As is well-known, 
by using CMB observations, we cannot in principle 
see the difference between those modes although 
they can give a severe constraint on the sum of
the size (fraction) of such isocurvature modes.
However, once it is confirmed that CDM/baryon isocurvature perturbation should contribute 
to density fluctuations in the Universe even if the size is small, it would be very important to 
differentiate those modes  since they can  give invaluable information 
on the nature of dark matter/generation mechanism of baryon asymmetry of the Universe. 

We showed that 21 cm fluctuation survey, which effectively probes density fluctuations 
of baryon, can in principle differentiate CDM and baryon isocurvature modes 
as shown in Figs.~\ref{fig:pk_21_1} and \ref{fig:pk_21_2}. 
To see this issue in some quantitative manner, we made a Fisher matrix analysis 
and discussed to what extent 21 cm survey can remove the complete degeneracy 
between the fractions of CDM and baryon isocurvature modes
which resides in CMB. When isocurvature 
fluctuations have scale-invariant primordial spectra, 
it seems very difficult to see the difference  even with the FFT telescope. 
However, if isocurvature modes have very blue-tilted spectra, which 
is claimed to be favored by observations \cite{Keskitalo:2006qv,
Beltran:2004uv,Beltran:2005gr,Bean:2006qz,Sollom:2009vd,Li:2010yb},
 and is predicted in some axion model \cite{Kasuya:2009up}, 
it would be possible to distinguish CDM and baryon isocurvature modes, 
which we have explicitly shown in Fig.~\ref{fig:pink_rc_rb}.

The adiabaticity of density fluctuations can give 
important information on various aspects of the Universe.  In particular, 
in light that CDM/baryon isocurvature fluctuations might be generated in well-motivated 
scenarios of CDM and baryogenesis like axion models and Affleck-Dine baryogenesis, 
the differentiation of these two modes, which 
has been shown to be possible by using 21 cm fluctuation survey in this paper,
may well give important implications for understanding
the nature of CDM and baryogenesis scenario.

\section*{Acknowledgments}

T. S. would like to thank the Japan Society for the Promotion
of Science for financial support.
This work is supported by Grant-in-Aid for Scientific research from
the Ministry of Education, Science, Sports, and Culture, Japan,
Nos.~14102004 (M.K.) and 21111006 (M.K.) and 19740145 (T.T.), and also by World Premier
International Research Center Initiative (WPI Initiative), MEXT, Japan.

\appendix

%
%
%

\end{document}